\documentclass{aastex631}
\usepackage{amsmath}

\shorttitle{Multiwavelength spectroscopic Sun-as-a-star analysis}
\shortauthors{Otsu {\&} Asai}

\graphicspath{{./}{figures/}}

\begin{document}

\title{Multiwavelength Sun-as-a-star Analysis of the M8.7 Flare on 2022 October 2 \\Using H$\alpha$ and EUV Spectra Taken by SMART/SDDI and SDO/EVE}

\author{Takato Otsu}
\affiliation{Astronomical Observatory, Kyoto University, Sakyo, Kyoto, Japan}

\author{Ayumi Asai}
\affiliation{Astronomical Observatory, Kyoto University, Sakyo, Kyoto, Japan}
 
\begin{abstract}
This paper presents a multiwavelength Sun-as-a-star analysis of the M8.7 flare on 2022 October 2, which were associated with a filament eruption and the following coronal mass ejection.
The Sun-as-a-star analysis was performed using H$\alpha$ data taken by Solar Dynamics Doppler Imager on board the Solar Magnetic Activity Research Telescope at Hida Observatory, Kyoto University and full-disk integrated extreme ultraviolet (EUV) spectra taken by the Extreme ultraviolet Variability Experiment (EVE) on board the Solar Dynamics Observatory. 
The Sun-as-a-star H$\alpha$ spectra showed blueshifted absorption corresponding to the filament eruption.
Furthermore, the EVE O {\sc v} 629.7 {\AA} spectra showed blueshifted brightening, which can also be attributed to the filament eruption. Even when the blueshifted absorption became almost invisible in the Sun-as-a-star H$\alpha$ spectra, the O {\sc v} blueshifted brightening up to $-400$ km s$^{-1}$ was still clearly visible. This result indicates that even when the shifted components--which are expected to originate from stellar eruptions--become almost invisible in the spatially integrated stellar H$\alpha$ spectra, the erupting materials may still be present and observable in EUV spectra.
Additionally, the Sun-as-a-star H$\alpha$ and O {\sc v} spectra exhibited redshifted absorption and brightening, respectively, during the decay phase of the flare.
These components probably originate from the post-flare loops, providing clues for the multi-temperature nature of the post-flare loops in the spatially integrated observation. Our Sun-as-a-star results suggest that the combination of H$\alpha$ and EUV lines allows the investigation of the multi-temperature structure and temporal development of stellar active phenomena even in spatially integrated spectra.
\end{abstract}

\keywords{Solar flares (1496); Stellar flares (1603); Solar filament eruptions (1981);
Solar prominences (1519); Solar coronal mass ejections (310); Stellar coronal mass ejections (1881); Solar chromosphere (1479)}

\section{Introduction} \label{sec:intro}
Solar flares are known as sudden brightening in the solar atmosphere. Typical solar flares are accompanied by various activities such as filament (or prominence) eruptions, coronal mass ejections (CMEs), chromospheric evaporation/condensation, and post-flare loops, as expected in the standard flare model \citep[for reviews,][]{ShibataMagara2011,Benz2017LRSP...14....2B}.
In the case of the Sun, most solar active phenomena exhibit multi-temperature structures.
For example, solar filament/prominence eruptions are often observed in H$\alpha$ and extreme ultraviolet (EUV) lines, simultaneously \citep[e.g.,][]{GutierrezETAL2021}. 
Furthermore, the observable wavelengths for filament/prominence eruptions change from cooler (e.g., H$\alpha$ $\sim10^4$ K) to hotter (e.g., UV, EUV $\sim10^5$ K) lines \citep[e.g.,][]{Fontenla1989,Chifor2006A&A...458..965C}.
In the standard flare model, the energy released by magnetic reconnection at the corona flows down to the chromosphere along the magnetic loops.
Subsequently, the chromosphere is heated, and hot and dense materials rise up and fill the magnetic loops (chromospheric evaporation).
This hot and dense loops are observed in soft X-ray ($\sim10^{7}$ K).
And as a result of a cooling process, loops become cooler and observable from EUV to H$\alpha$.
Such loop structures in soft X-ray to H$\alpha$ are usually called 'post-flare loops' \citep{Bruzek1964ApJ...140..746B,Kamio2003SoPh}.
In post-flare loops, a catastrophic cooling process induced by thermal instability generated dense blobs \citep[e.g.,][]{Ruan2021ApJ...920L..15R}, which are observed as downflows towards the solar surface along the post-flare loops in EUV and/or H$\alpha$ \citep[e.g.,][]{Antolin2022FrASS...920116A}.

Similar to solar flares, stellar flares are known as sudden brightening on stars; especially, those emitting energies more than $10^{33}$ erg are called superflares \citep[e.g.,][]{MaeharaETAL2012,OkamotoETAL2021}.
Stellar superflares may be accompanied by substantially larger CMEs than those observed on the Sun, and such super CMEs are considered to greatly affect the exoplanets around the host stars \citep[e.g.,][]{AirapetianETAL2020, CliverETAL2022}.
Hence, superflares are actively investigated, not only from a physical viewpoint but also with respect to the habitability of exoplanets.
In fact, recent observations have revealed various signatures associated with stellar (super-) flares, suggesting that stellar flares are accompanied by various activities like solar flares \citep[e.g.,][]{Vida2019A&A...623A..49V, MuhekiETAL2020, MaeharaETAL2021,VeronigETAL2021,NamekataETAL2022a,Wu2022ApJ...928..180W,NamizakiETAL2023ApJ, InoueETAL2023ApJ,Notsu2023arXiv231002450N,Namekata2023arXiv231107380N}.
\citet{MaeharaETAL2021} analyzed temporally resolved H$\alpha$ spectra of stellar flares on an M-type star and found a blueshifted excess emission, which might be arising from a prominence eruption.
Additionally, \citet{InoueETAL2023ApJ} conducted a similar analysis of a superflare on an RS CVn-type star and found a blueshifted component with very high speed--which exceeded the escape velocity of the target star-- suggesting the occurrence of a gigantic CME.
\citet{NamizakiETAL2023ApJ} analyzed a superflare on the same M-type star as in \citet{MaeharaETAL2021}, and found redshifted excess emission with long duration in H$\alpha$ spectra which was exhibited not only during the flare impulsive phase but also during the flare decay phase. They proposed that the redshifted emission observed during the flare decay phase came from downflows towards the stellar surface along the post-flare loops.
These observations strongly suggest that stellar flares are not just a brightening of stars but are dynamic phenomena with various associated activities like solar flares.

Identifying what occurs on stellar surfaces is, however, difficult because the surfaces of distant stars cannot be resolved with current observational instruments.
Hence, some studies have utilized solar data with spatial resolution to interpret stellar data without spatial resolution by performing Sun-as-a-star analyses \citep[e.g.,][]{ToriumiETAL2020,NamekataETAL2022a,NamekataETAL2022c,Namekata2023ApJ...945..147N,OtsuETAL2022,XuETAL2022,LuETAL2023}.
In Sun-as-a-star analyses, solar data are integrated spatially to be compared with stellar data. Owing to the spatial integration, the information regarding motions of materials on the plane of the sky are lost.
Thus, Sun-as-a-star analyses with spectral data--which provide information on the line-of-sight dynamics-- are useful for investigating the flare dynamics and plasma motions.
For example, \citet{NamekataETAL2022a} performed Sun-as-a-star analyses of H$\alpha$ spectra for two flares with a filament eruption/surge and compared the resulting spatially integrated H$\alpha$ spectra with stellar data. 
Based on the similarity between the solar and stellar data, they concluded that the observed stellar flare was also accompanied by a filament eruption. Later, \citet{OtsuETAL2022} analyzed H$\alpha$ spectra for nine solar active events, and demonstrated that the diversity in solar activities, such as filament/prominence eruptions, chromospheric condensations and downflows along post-flare loops, can be confirmed even in the spatially integrated H$\alpha$ spectra. The Doppler signals in Sun-as-a-star spectra are also analyzed using the EUV spectral lines taken by the Extreme ultraviolet Variability Experiment \citep[EVE;][]{WoodsETAL2012} onboard the Solar Dynamic Observatory \citep[SDO;][]{PesnellETAL2012}.
SDO/EVE has been mainly used for analyzing a light curve of each EUV line \citep[e.g.,][]{HarraETAL2016,Nishimoto2020ApJ}.
Furthermore, recent many studies have firmly established the capability of Doppler analysis for solar active phenomena by SDO/EVE \citep{Hudson2011SoPh,Chamberlin2016SoPh,Cheng2019ApJ,XuETAL2022,Hudson2022MNRAS,Fitzpatrick2023SoPh,LuETAL2023}.
Recently, \citet{XuETAL2022} performed a Sun-as-a-star analysis of the X1.0 flare observed on 2021 October 28 using EUV spectral lines taken by SDO/EVE. 
They found a blueshifted component in pre-flare subtracted spectra of some EUV spectral lines, such as O {\sc v} 629.7 {\AA}, which were attributed to a filament eruption associated with the X1.0 flare. 
Later, \citet{LuETAL2023} analyzed eight flares associated with CMEs using the almost same method with that in \citet{XuETAL2022}, and they also obtained blueshifted components related to the CMEs in the Sun-as-a-star EUV spectra. 

As introduced above, so far, spectroscopic Sun-as-a-star analyses of solar active phenomena have been performed mainly using either H$\alpha$ or EUV spectra.
However, Sun-as-a-star analysis simultaneously using H$\alpha$ and EUV-- which have different formation temperatures-- have not yet been performed. Solar activities, such as filament/prominence eruptions and post-flare loops,
generally show multi-temperature structures. 
Thus, lines in a single temperature range are insufficient to grasp the whole dynamics of solar active phenomena, which should also be applicable to the stellar activities.
Therefore, Sun-as-a-star analyses simultaneously using multi-line spectra with different formation temperature are required for more detailed investigation of the flare dynamics and plasma motions on stars.

In this paper, we present the Sun-as-a-star analysis of an M8.7 flare observed simultaneously with H$\alpha$ and EUV spectra on 2022 October 2.
Herein, such simultaneous analysis using H$\alpha$ and EUV spectra has been performed for the first time.
The observational data is presented in Section \ref{Obs}.
The methods for performing Sun-as-a-star analyses are detailed in Section \ref{Methods}. 
In Section \ref{Re}, we report the results and their correspondence with active phenomena in spatially resolved images. Further, we provide discussions in Section \ref{Dis}. 
Finally, we summarize the present paper in Section \ref{Sum}.

%============================================================
\section{Observations}\label{Obs}
In this paper, we used two instruments for performing a Sun-as-a-star analysis: Solar Dynamics Doppler Imager \citep[SDDI;][]{IchimotoETAL2017} on board the Solar Magnetic Activity Research Telescope \citep[SMART;][]{UenoETAL2004} at Hida Observatory, Kyoto University for H$\alpha$ line spectra and SDO/EVE for EUV spectra. The SDDI takes full-disk solar images at 73 wavelengths from the H$\alpha$ $-9.0$ {\AA} to H$\alpha$ $+ 9.0$ {\AA} with a spectral sampling of 0.25 {\AA}. 
The time cadence and pixel size are 12 s and 1$^{\prime\prime}$.23, respectively.
The EVE takes solar EUV irradiance with a spectral resolution of $\sim1$ {\AA} (sampling is $\sim0.2$ {\AA}). 
%and a time cadence of $\sim10$ s. 
The Multiple EUV Grating Spectrographs-A/B (MEGS-A/B), originally installed in EVE,
could observe wavelength in the range of 50–370 {\AA} and 330–1066 {\AA}, respectively. 
However, MEGS-A experienced a CCD anomaly on 2014 May 26, 
and since then, the spectra at wavelengths shorter than 330 {\AA} are unavailable.
In this study, we used EVE Level 2B spectra (EVS\_L2B) with a time cadence of 1 min.
% For a higher signal-to-noise ratio, the data of MEGS-B was integrated over 1 minute. 
Additionally, We used imaging data taken by Atmospheric Imaging Assembly \citep[AIA;][]{LemenETAL2012AIA} on board the SDO for confirming the dynamics of a flare and eruption in EUV wavelengths. 

\subsection{Event overview}
\begin{figure}[htbp]
\centering
\includegraphics[width=18cm]
{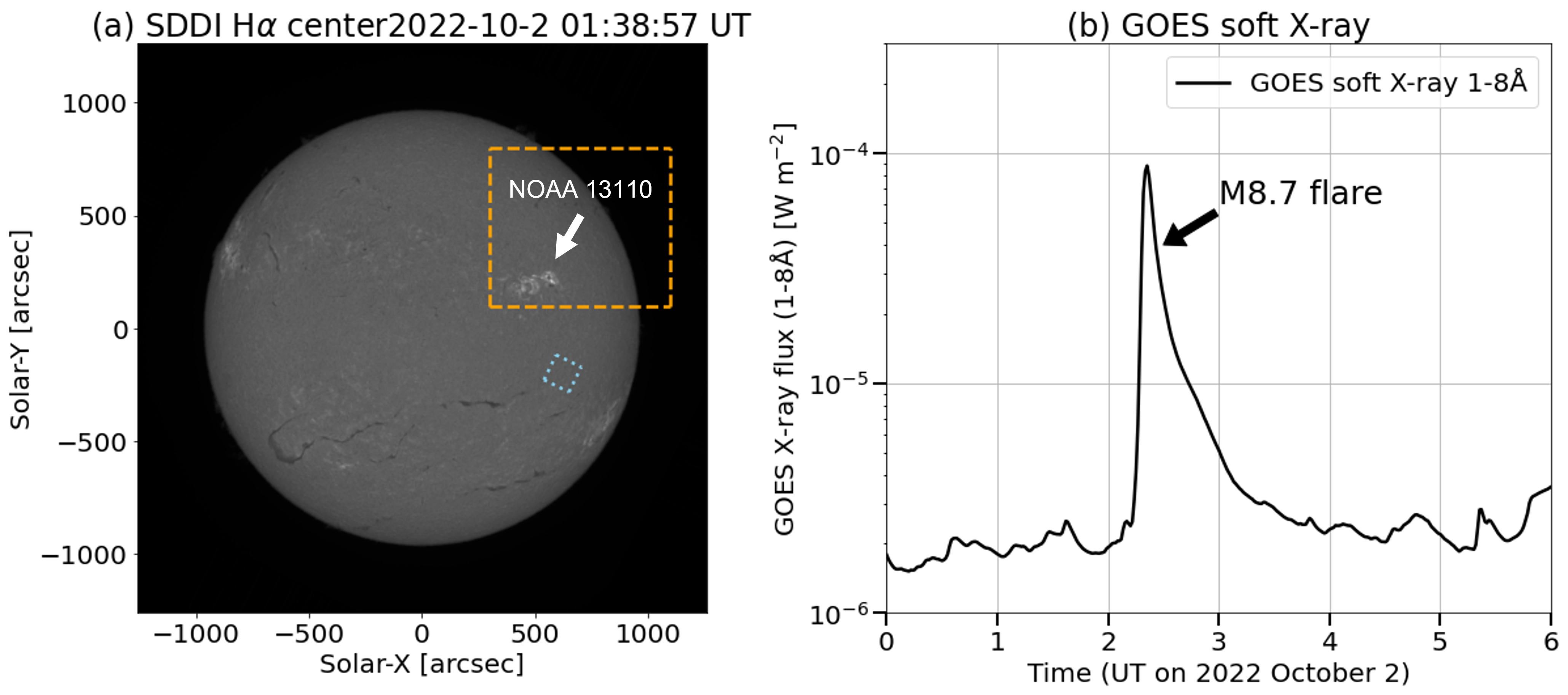}
\caption{(a) The solar full-disk image taken by SMART/SDDI at the H$\alpha$ line center observed on 2022 October 2 01:38:57 UT. Solar north and west are at the top and right, respectively. The white arrow indicates the target active region NOAA 13110. The orange dashed and sky blue dotted boxes correspond to the field of view in Figure \ref{Eruption} and the quiet region for calibration (see Section \ref{Methods}), respectively.
(b) GOES soft X-ray light curve between 00:00 UT and 06:00 UT on 2022 October 2 is shown as black solid line. The black arrow indicates the target M8.7 flare.}
\label{full}
\end{figure}

\begin{figure}[htbp]
\centering
\includegraphics[width=18cm]
{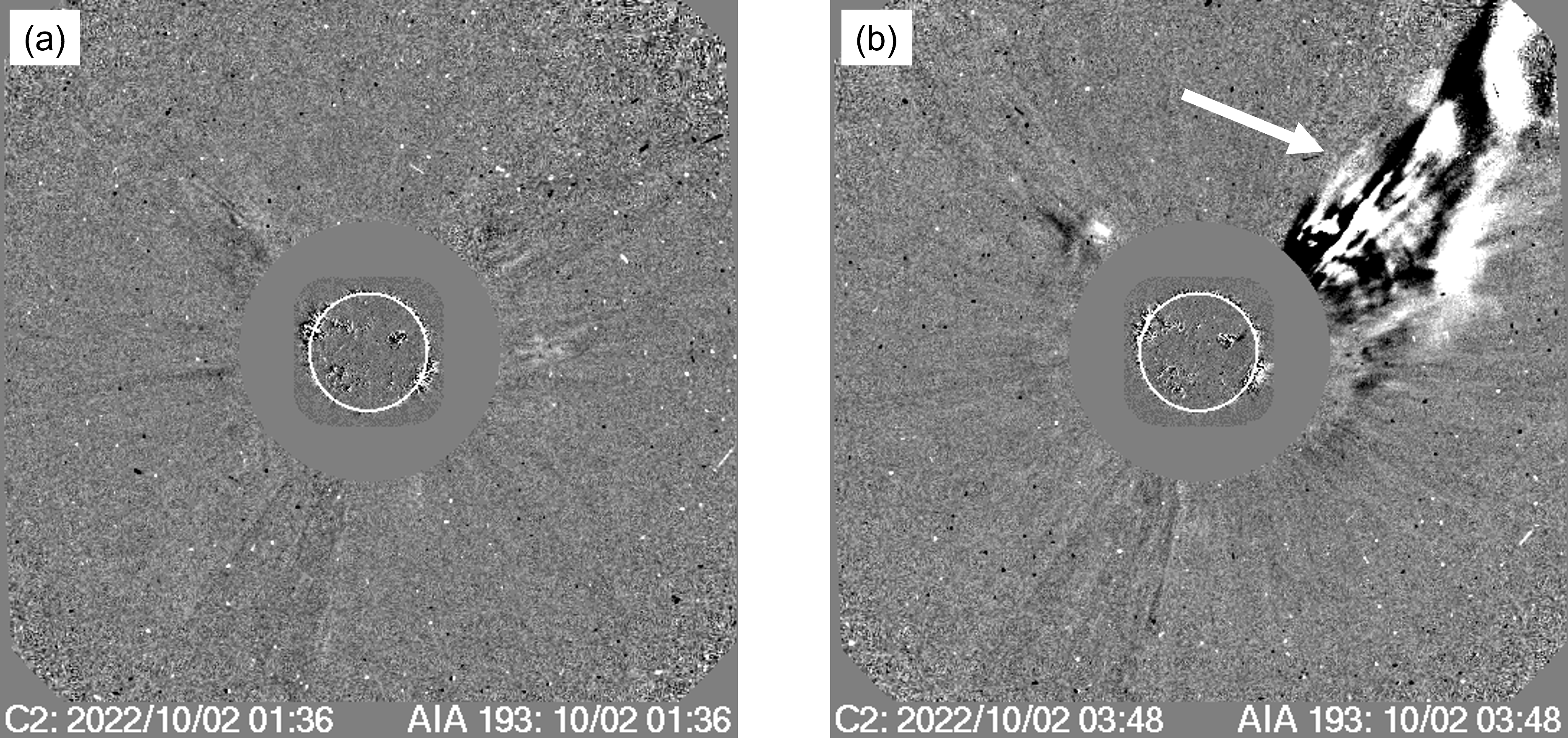}
\caption{The CME associated with the target M8.7 flare. LASCO C2 running difference images at 01:36 UT and 03:48 UT on 2022 October 2 are shown in panels (a) and (b), respectively. Additionally, an AIA 193 image is shown and solar limb is indicated as a white circle in each panel. Panel (a) corresponds to the pre-CME phase. In panel (b), the white arrow indicates the CME associated with the M8.7 flare. The images are obtained from \url{https://cdaw.gsfc.nasa.gov/}.}
\label{CME}
\end{figure}

\begin{figure}[htbp]
\centering
\includegraphics[width=14cm]
{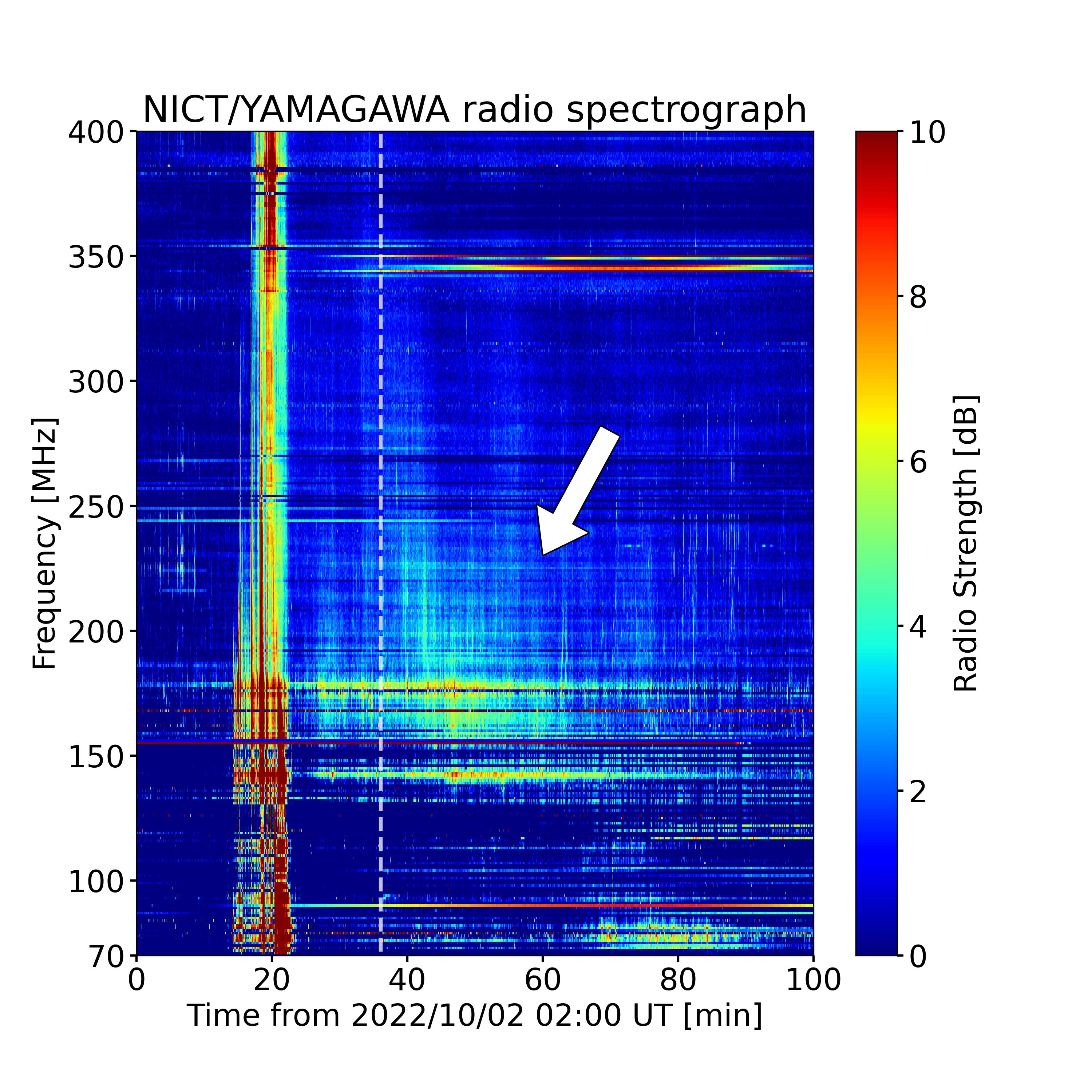}
\caption{The radio burst associated with the M8.7 flare. The radio dynamic spectrum taken by NICT/YAMAGAWA radio spectrograph is shown. 
The vertical dashed line around $t=36$ min indicates the CME appearance time in SOHO/LASCO C2 (02:36 UT)
The white arrow indicates the radio enhancement around the CME time.}
\label{Radio}
\end{figure}

The overview of the target event is as follows. 
The target event is the M8.7 flare that occurred on 2022 October 2 in the NOAA 13110
which was located on N17W39.
Figure \ref{full} (a) shows the full-disk image of the H$\alpha$ line center taken by the SDDI on 2022 October 2 01:38:57 UT. 
The target active region is indicated using the white arrow.
The flare started at 02:08 UT and peaked at 02:21 UT in GOES soft X-ray as indicated by the black arrow in Figure \ref{full} (b).
This flare was simultaneously observed by the SDDI and EVE, although the EVE missed a part of the impulsive phase from 02:07 UT to 02:20 UT. The SDDI and AIA took the images of the flare and the associated filament eruption (Figure \ref{Eruption}).
The related CME was observed by the Large Angle and Spectrometric Coronagraph \citep[LASCO;][]{BruecknerETAL1995} on board the Solar and Heliospheric Observatory \citep[SOHO;][]{Domingo1995SoPh..162....1D} as shown in Figure \ref{CME}.
Figure \ref{Radio} shows the radio dynamic spectrum taken by YAMAGAWA of National Institute of Information and Communications Technology
(NICT) \citep{Iwai2017EP&S...69...95I} in the frequency range 70-9000 MHz. The vertical dashed line in Figure \ref{Radio} indicates the CME appearance time in SOHO/LASCO C2 (02:36 UT).
As indicated by the white arrow in Figure \ref{Radio}, the radio enhancement can be confirmed around the CME time but it is not so strong. This event was also associated with various other phenomena related to the flare, such as chromospheric condensation at the flare ribbons (Figure \ref{RA}), post-flare loops with cool downflows, and dimming which colud be related to the CME (Figure \ref{CRs}).
We will describe details of these phenomena in the following subsections.

\subsection{Filament eruption \label{Re_event}}
\begin{figure}[htbp]
\centering
\includegraphics[width=11cm]
{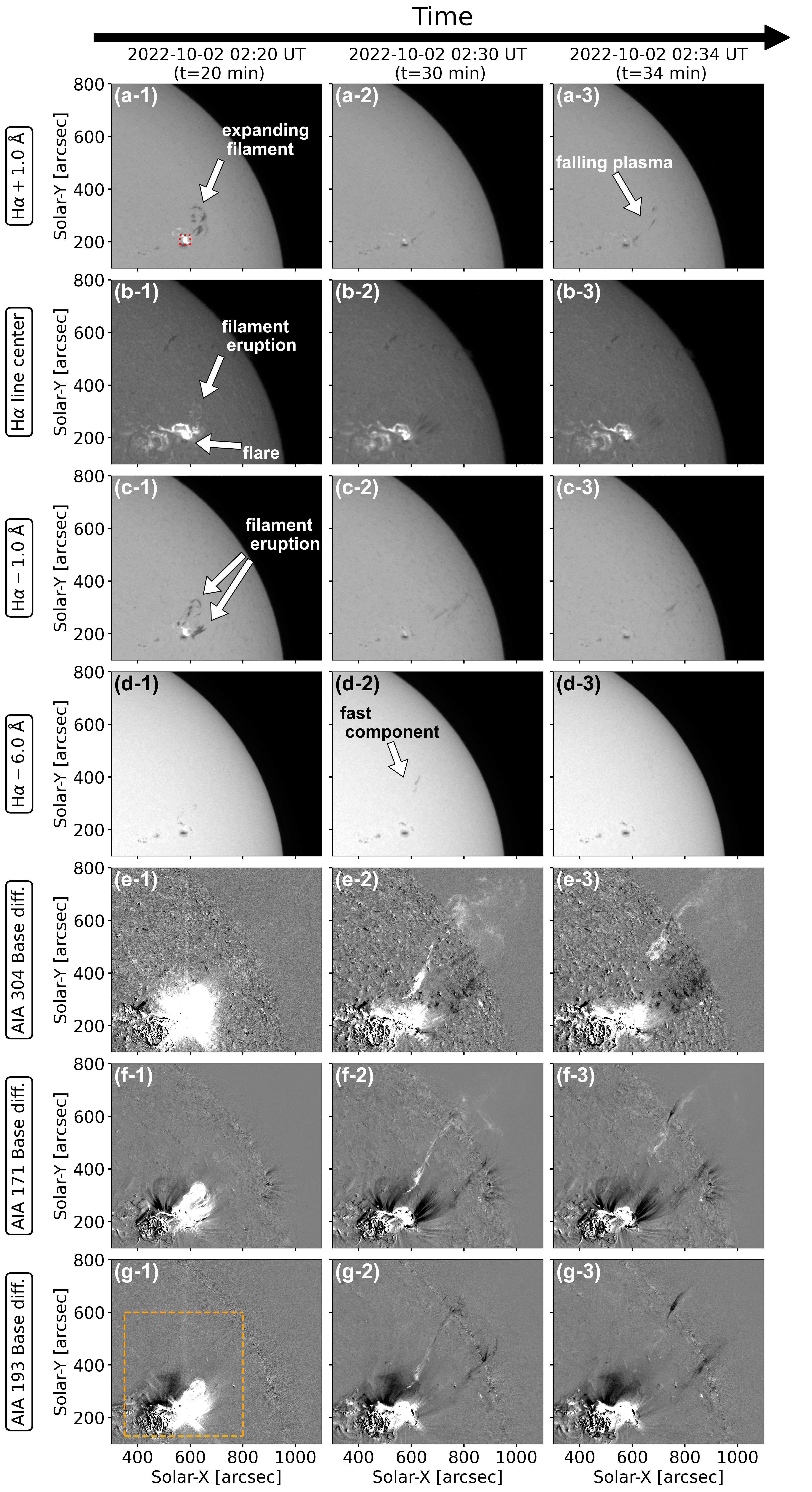}
\caption{Time development of the target event in H$\alpha$ spectral images and EUV images taken by SMART/SDDI and SDO/AIA, respectively. From left to right, images at 02:20 UT (t=20 min), 02:30 UT (t=30 min) and 02:34 UT (t=34 min) are shown. The time $t$ is measured from 02:00 UT. In the top four rows, images of H$\alpha+1.0$ {\AA} (a-1)-(a-3), H$\alpha$ line center (b-1)-(b-3), H$\alpha-1.0$ {\AA} (c-1)-(c-3) and H$\alpha-6.0$ {\AA} (d-1)-(d-3) are shown. In the bottom three rows, base-difference images of AIA 304 {\AA} (e-1)-(e-3), 171 {\AA} (f-1)-(f-3) and 193 {\AA} (g-1)-(g-3) are shown. The base of these images is the data at 02:00 UT for each channel. Some notable points are indicated by white arrows.
The field of view of the all panels corresponds to the orange dashed box in the panel (a) of Figure \ref{full}.
The red dotted box in (a-1) and orange dashed box in (g-1) correspond to the field of view in Figures \ref{RA} and \ref{CRs}, respectively.}
\label{Eruption}
\end{figure}

Figure \ref{Eruption} shows the temporal development of the M8.7 flare in the H$\alpha$ spectral images taken by SMART/SDDI, and in base-difference images of EUV channels taken by SDO/AIA. The H$\alpha$ line center images show brightening from the flare ribbons of the M8.7 flare (Figure \ref{Eruption} (b-1)-(b-3)). 
Additionally, the filament eruption can be confirmed as a bright material in the H$\alpha$ line center image at an early phase of the flare (Figure \ref{Eruption} (b-1)).
Such bright features of erupting filaments suggest heating of materials during the eruption and have been reported by previous studies \citep[e.g.,][]{Ding2003ApJ...598..683D}.
The images of H$\alpha$ blue wing demonstrate erupting filament as dark features moving toward north-west direction (Figure \ref{Eruption} (c-1)-(c-3), (d-2)).
The line-of-sight velocity of the erupting filament on the solar disk exceeded $\approx -300$ km s$^{-1}$, corresponding to approximately H$\alpha-6.0$ {\AA} (Figure \ref{Eruption} (d-2)).
Note that the some parts of the erupting material can still be identified beyond the limb.
However, radiation from these components becomes fainter and fainter as they move outward.
The erupting filament can be also confirmed as dark features in the images of H$\alpha$ red wing (Figure \ref{Eruption} (a-1)).
This redshifted component corresponds to materials moving away from the Earth
and may arise from the expanding and/or helical motion of the erupting filament.
After the eruption, dark features moving back to the solar surface can be confirmed in the red wing images (Figure \ref{Eruption} (a-3)), ascribing to the partial drainage of the erupted filament. 
Similar to the H$\alpha$ line, the flare brightening was evident in the base-difference images of the AIA 304, 171, and 193 {\AA} channels (Figure \ref{Eruption} (e)-(g)).
The associated filament eruption can also be seen in these AIA channels.
%in the all channels, although its contrast to the solar surface is weaker in hotter channels. 
At the initial phase of the eruption, the filament eruption is clearly evident in the H$\alpha$ and the AIA images (Figure \ref{Eruption} (a-1)-(c-1), (e-1)-(g-1) $t=20$ minutes).
On the other hand,
at the later phase, the erupting filament can be seen in the AIA images even though it was almost invisible in the H$\alpha$ spectral image (Figure \ref{Eruption} (a-2)-(d-2), (e-2)-(g-2) $t=30$ min; (a-3)-(d-3), (e-3)-(g-3)) $t=34$ min).
The filament eruption possibly developed to the core of the CME taken by SOHO/LASCO (Figure \ref{CME} (b)).
%-----------------------------
\subsection{Chromospheric condensation}
\begin{figure}[htbp]
\centering
\includegraphics[width=16cm]
{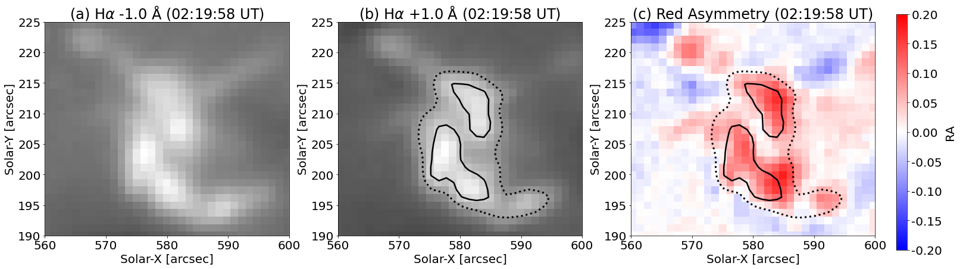}
\caption{Zoomed up images of the flare ribbons and red asymmetry. Images of H$\alpha-1.0$ {\AA} and H$\alpha+1.0$ {\AA} taken by SMART/SDDI and distribution of red asymmetry (RA; defined in Equation \ref{RAEQ}) on 02:19:58 UT are shown in panels (a), (b) and (c), respectively.
The field of view of (a)-(c) corresponds to the orange dashed box in the panel (a-1) of Figure \ref{Eruption}.
The solid and dotted lines in (b) are 80{\%} and 60{\%} of the maximum intensity in the image of (b), respectively. The same contours with those in (b) are plotted in (c).}
\label{RA}
\end{figure}

Figure \ref{RA} shows zoomed-in images around the two ribbons of the flare (the red dotted box in Figure \ref{Eruption} (a-1) and Figure \ref{CRs} (a-1)) in H$\alpha\pm1.0$ {\AA} and red asymmetry (RA) during the impulsive phase of the flare (02:19:58 UT).
The RA is defined by the following equation;

\begin{equation}
    RA(x,y,t)=\frac{I(x,y,\Delta\lambda=+1.0 \mathrm{\AA}, t)-I(x,y,\Delta\lambda=-1.0 \mathrm{\AA}, t)}
    {I(x,y,\Delta\lambda=+1.0 \mathrm{\AA},t)+I(x,y,\Delta\lambda=-1.0 \mathrm{\AA},t)},
    \label{RAEQ}
\end{equation}
where, $I(x,y,\Delta\lambda,t)$ is the intensity as a function of position $(x,y)$, $\Delta\lambda$ is the difference of wavelength from H$\alpha$ line center (6562.8 {\AA}), and $t$ is the time. 
In the panels (b) and (c) of Figure \ref{RA}, the contours corresponding to the 80\% and 60\% of the maximum intensity at H$\alpha+1.0$ {\AA} over the field of view in Figure \ref{RA} are shown as the solid and dotted lines, respectively.
The RA has a positive value when the intensity of H$\alpha$ red wing ($\Delta\lambda=+1.0$ {\AA}) is higher than that of H$\alpha$ blue wing  ($\Delta\lambda=-1.0$ {\AA}).
Thus, RA can be used as an indicator of red asymmetry \citep{AsaiETAL2012,NamekataETAL2022c}. The RA in panel (c) of Figure \ref{RA} shows a positive value on the two ribbons, which is usually attributed to downflows of the chromospheric condensation \citep[e.g.,][]{SvestkaETAL1962, IchimotoKurokawa1984, AsaiETAL2012}.
%----------------------------
\subsection{Post-flare loops and dimming \label{PFL_Dim}}
\begin{figure}[htbp]
\centering
\includegraphics[width=16cm]
{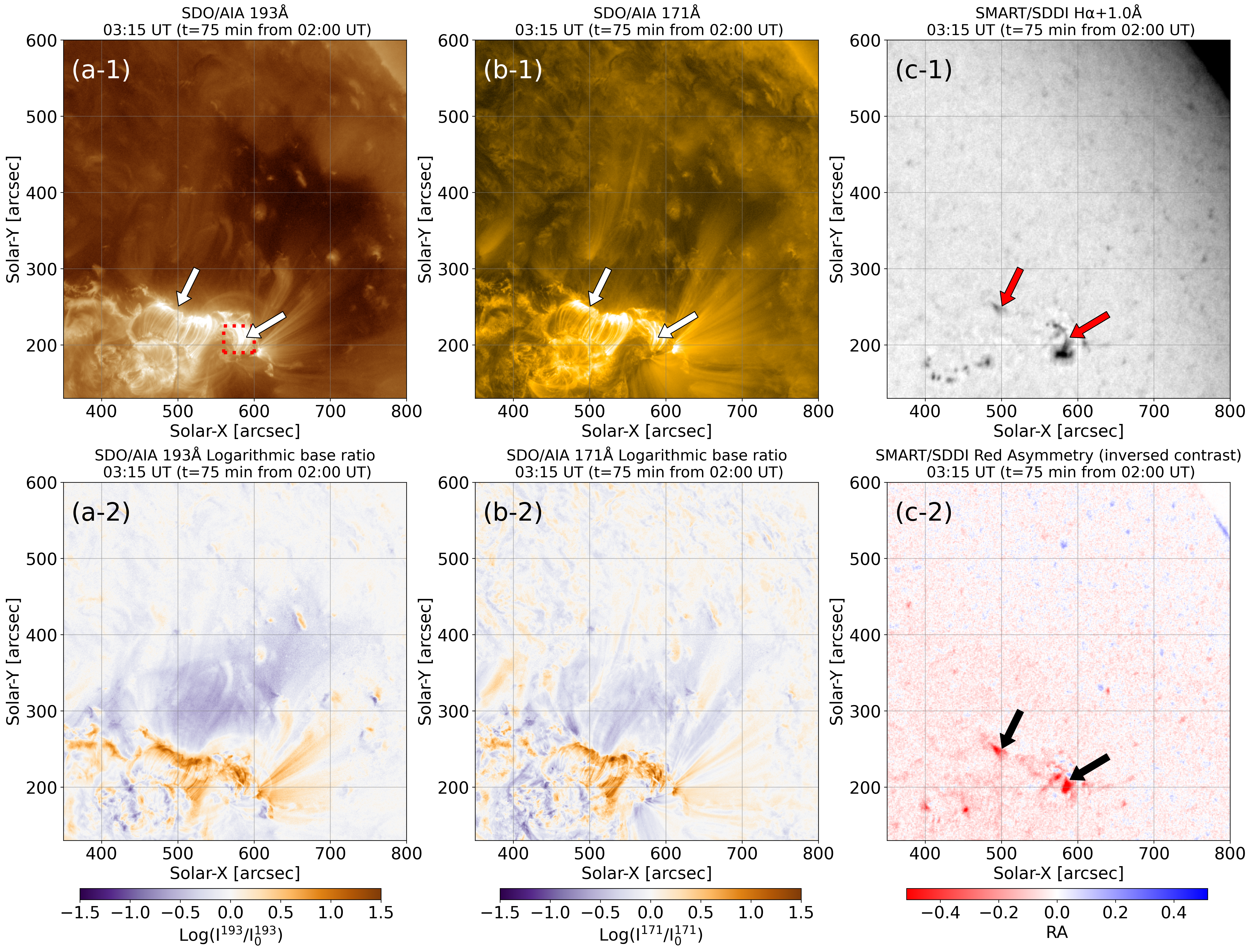}
\caption{Post-flare loops and dimming in AIA channels and downflows around the post-flare loops in H$\alpha$. Panels (a-1), (b-1) and (c-1) show images of AIA 193 {\AA}, 171 {\AA} and SDDI H$\alpha+1.0$ {\AA} on 03:15 UT, respectively. Panels (a-2) and (b-2) show distributions of logarithmic base ratio on 03:15 UT for AIA 193 {\AA} and 171 {\AA}, respectively. Panel (c-2) demonstrates distribution of red asymmetry similar to panel (c) of Figure \ref{RA} but indicated as inverse contrast to show redshifted dark downflows in red color. 
The field of view for all panels corresponds to the red dotted box in the panel (g-1) of Figure \ref{Eruption}.
The arrows in (a-1) and (b-1) indicate post-flare loops, on the other hand, while those in (c-1) and (c-2) indicate downflows around the post-flare loops in (a-1) and (b-1).}
\label{CRs}
\end{figure}

Figure \ref{CRs} shows zoomed-in images around the flare region (the orange dashed box in Figure \ref{Eruption} (g-1)) in the decay phase of the flare (03:15 UT: about 50 minutes after the GOES peak time 02:21 UT). 
Panels (a-1) and (b-1) in Figure \ref{CRs} show the images of AIA 193 and 171 {\AA}, respectively.
Panels (a-2) and (b-2) in Figure \ref{CRs} show the logarithmic base ratio images  of AIA 193 and 171 {\AA} \citep{Dissauer2018ApJ...855..137D}, which enhance the brightening and darkening as orange and purple colors, respectively. 
Panels (c-1) and (c-2)  in Figure \ref{CRs} show the images of  H$\alpha$ red wing and RA. Note that the RA is shown as inversed color map compared
to that in Figure \ref{RA} (c) in order to show the dark downflows in red color.

\begin{figure}[htbp]
\centering
\includegraphics[width=17cm]
{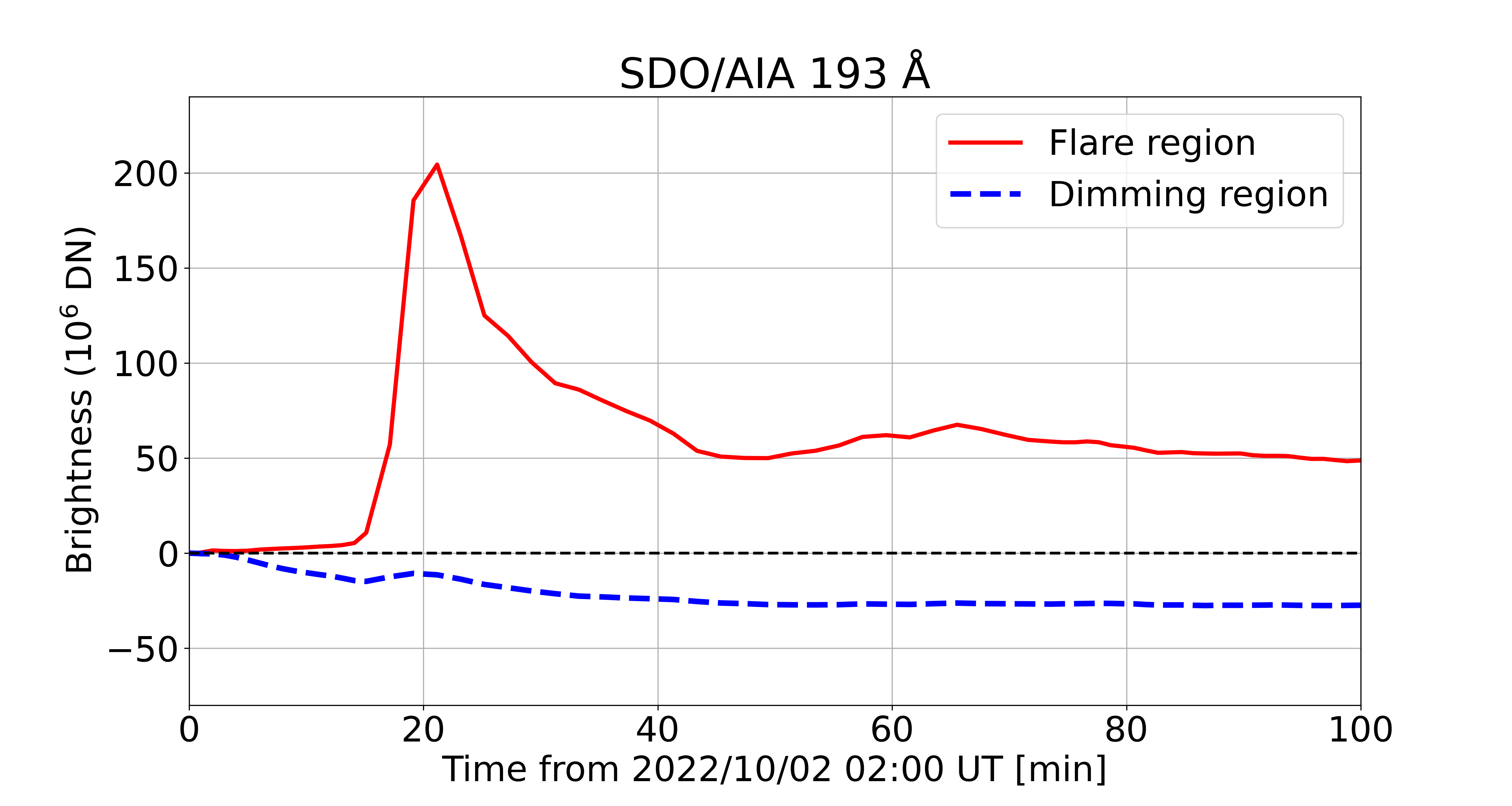}
\caption{AIA 193 {\AA} light curves for flare and dimming regions. The AIA 193 {\AA} light curves integrated in the flare and dimming regions (see text) are plotted as red solid and blue dashed curves, respectively. The horizontal black dashed line indicates the pre-event level.}
\label{LC_df}
\end{figure}

As indicated by the white arrows in Figure \ref{CRs} (a-1) and (b-1), post-flare loops can be confirmed in AIA 171 and 193 {\AA} channels.
Additionally, as indicated by the red and black arrows in panels (c-1) and (c-2) in Figure \ref{CRs}, the downflows in H$\alpha$ spectra are almost co-located with the post-flare loops in AIA 171 and 193 {\AA}.
This co-location implies that the downflows in the H$\alpha$ images correspond to cooled material falling along the post-flare loops.
Around the bright post-flare loops, dimming regions are confirmed in AIA 193 and 171 {\AA} (Figure \ref{CRs} (a-2) and (b-2)). 
Although such dimming can also occur in low-temperature channels such as AIA 171 {\AA} due to the heating of materials \citep[e.g.,][]{Mason2014ApJ...789...61M}, the dimming in this event could be related to the CME (Figure \ref{CME}) since the dimming is clearly visible in hot AIA 193 {\AA} image.
Figure \ref{LC_df} shows the AIA 193 {\AA} light curves of the flare and dimming regions.
These light curves were obtained mainly by following the process in \citet{VeronigETAL2021}.
In each time step, we defined the flare and dimming regions,
where logarithmic base ratio is larger than $0.18$ and lower than $-0.15$, respectively, within the field of view of Figure \ref{CRs}.
These two light curves were obtained by summing up the pixel values in each region subtracted by the pre-event counts. 
The light curve for the flare region shows two distinct increases.
The first increase from $t\approx15$ min corresponds to the impulsive phase of the flare (Figure \ref{Eruption} (g-1)), whereas
the secondary increase from $t\approx50$ min represents the evolving post-flare loops (Figure \ref{CRs} (a-1)-(a-2)).
The light curve for the dimming region shows decrease from $t\approx5$ min.
Notably, the amount of decrease for the dimming region is smaller than that of increase for the flare region during the almost all phase.

%============================================================
\section{Methods of Sun-as-a-star Analyses\label{Methods}}
\subsection{Sun-as-a-star analysis of H$\alpha$ spectral images}
\subsubsection{Masked region including the flare and filament eruption}
The filament in this event moves in the wide field of view during its eruption ($\sim R_{sun}$ (the solar radius)).
The spatial integration over a simple square region of $\sim R_{sun}\times R_{sun}$ provides no valuable signature due to noises such as the Earth’s atmospheric fluctuations.
Thus, we defined a masked region including the flare and filament eruption,
and performed integration over it.
The noisy data at limb and relatively noiseless data outside the solar disk were manually removed and included, respectively.
The masked region was obtained using the following process.
First, we took out a north-west quarter area of the solar full disk in the original observation of the SDDI, where we conduct the next step.
Second, we defined the H$\alpha$ spectral intensity $I(x, y, \Delta\lambda, t)$ as a function of position $(x,y)$, difference in wavelength from the H$\alpha$ line center $\Delta\lambda$ and time $t$. Subsequently, we calculated normalized spectra using a reference spectra $I_{QR}(\Delta\lambda, t)=<I(x, y, \Delta\lambda, t>_{QR}$,
where $<>_{QR}$ means spatial average over the quiet region (skyblue dotted region in Figure \ref{full} (a));
\begin{equation}
    \bar{I}(x,y,\Delta\lambda, t)=\frac{I(x, y, \Delta\lambda, t)}{I_{QR}(\Delta\lambda, t)} \times I_{QR}(\Delta\lambda, t_0),
\end{equation}
where $t_0=$ 01:39 UT is the start time of the data. Using $\bar{I}(x, y, \Delta\lambda, t)$, we obtained the change in spectra from the pre-event state;

\begin{equation}
    \bar{I}^{pre}(x,y,\Delta\lambda)=<\bar{I}(x, y, \Delta\lambda, t)>_{t=t_0}^{t=t_{pre}}
\end{equation}

\begin{equation}
    \Delta\bar{I}(x, y, \Delta\lambda, t)=\bar{I}(x,y,\Delta\lambda, t)-\bar{I}^{pre}(x,y,\Delta\lambda),
\end{equation}
where $<>_{t=t_0}^{t=t_{pre}}$ means the time average, which was taken between 01:39 UT and 01:51 UT.
Further, the standard deviation of pre-event data $\sigma (x,y,\Delta\lambda)$ was obtained as follows:
\begin{equation}
    \sigma(x, y, \Delta\lambda)=\sqrt{<(\Delta\bar{I}(x,y,\Delta\lambda, t))^2>_{t=t_0}^{t=t_{pre}}}.
\end{equation}

A masked region was obtained at each wavelength and time as the binary data $mask(x, y,\Delta\lambda, t)$;
\begin{equation}
    mask(x,y,\Delta\lambda, t)=\begin{cases}
        1& \text{if $|\Delta\bar{I}(x,y,\Delta\lambda,t)| > 3\sigma(x,y,\Delta\lambda)$}\\
        0& \text{else}
    \end{cases}
\end{equation}
To remove spot noses and fill small gaps, morphological operators (IDL routines
dilute.pro and erode.pro) were applied to each mask. 
Similar process has also been applied in previous studies using data from the SMART/SDDI \citep{SekiETAL2017,SekiETAL2019a}.
Finally, we obtained the whole masked region as the binary data $Mask(x,y)$ by summing $mask(x, y, \Delta\lambda, t)$:

\begin{equation}\label{Mask}
    Mask(x, y)=\begin{cases}
        1& \text{if $\displaystyle\sum_{t}{\sum_{\Delta\lambda}{mask(x,y,\Delta\lambda, t)>0}}$}\\
        0& \text{else}
    \end{cases}
\end{equation}
where the sums with regard to wavelength $\Delta\lambda$ and time $t$ were performed in the ranges $[-9.0\mathrm{\AA},+9.0\mathrm{\AA}]$ and [02:00 UT, 03:30 UT], respectively.

\subsubsection{Pre-event subtracted H$\alpha$ spectra normalized by full-disk integrated continuum\label{Ha_Sun_as_a_star}}
The Sun-as-a-star analysis of H$\alpha$ spectral images taken by the SDDI is mainly based on the method of \citet{OtsuETAL2022} \citep[see also][]{NamekataETAL2022a,NamekataETAL2022c}.
A brief outline of the method used for obtaining the Sun-as-a-star H$\alpha$ spectra is as follows (see \cite{OtsuETAL2022} for details.).

\begin{itemize}
    \item [1.] Spatial integration of H$\alpha$ spectra was performed in a target region along with its normalization by using continuum of each integrated spectrum and data from quiet regions (blue dotted square in Figure \ref{full} (a)). The H$\alpha$ spectra at the positions $\{(x,y)\}$ with $Mask(x, y)=1$ were used for the spatial integration. 

    \item [2.] Pre-event subtraction was performed. The pre-event data for this subtraction was calculated by time averaging of the spatially integrated spectra between 01:39 UT and 02:00 UT.

    \item [3.] Pre-event subtracted spectra were normalized using a full-disk integrated continuum (irradiance at continuum).
\end{itemize}
The resulting normalized pre-event-subtracted H$\alpha$ spectra
$\Delta S_{H\alpha}(t,\lambda)$ represent the ratios of the spectral changes arising from the target event (e.g., the flare and filament eruption) to the solar irradiance.
Assuming that there are no other variations, except those in the masked region, $\Delta S_{H\alpha}(t,\lambda)$
approximately equals to the full-disk integrated pre-event subtracted spectra \citep{NamekataETAL2022a,NamekataETAL2022c,OtsuETAL2022}.
Hereafter, we call $\Delta S_{H\alpha}(t,\lambda)$ a Sun-as-a-star H$\alpha$ spectrum, omitting 'pre-event-subtracted'.
Additionally, we calculated the differenced H$\alpha$ equivalent width to obtain the light curve of the H$\alpha$ line: $\Delta EW(t)=\int^{\mathrm{H\alpha}+\Delta\lambda}_{\mathrm{H\alpha}-\Delta\lambda}\Delta S_{H\alpha}(t,\lambda)d\lambda$ where $\Delta\lambda$ was set as $6.0$ {\AA} to include the whole spectral variations (Figure \ref{LC} (b)). 

\subsection{Sun-as-a-star analysis of EUV}\label{EUV_SAS}
In this study, we focused on two spectral lines from the EVE MEGS-B lines; O {\sc v} 629.7 {\AA} and O {\sc vi} 1031.9 {\AA}.
These two lines have no serious line blend according to CHIANTI 10.1 \citep[][see also \citet{XuETAL2022}]{Dere1997AAS..125..149D, Dere2023ApJS..268...52D}. \citet{XuETAL2022} analyzed six lines of MEGS-B and reported that an obvious blueshifted component associated with plasma eruption in the O {\sc v} 629.7 {\AA} spectra.
Therefore, we focused on the O {\sc v} 629.7 {\AA} spectra to investigate the filament eruption in the transition region temperature.
The wavelength of O {\sc vi} is longer than that of the Lyman limit at 912 {\AA} and can be used for observation of distant stars, avoiding strong interstellar absorption \footnote{Note that O {\sc vi} 1031.9 {\AA} is often classified as a far ultraviolet (FUV) line} (\citet{Leitzinger2011A&A...536A..62L} shows examples of stellar observations in O {\sc vi} 1031.9 {\AA}.).
Thus, in this study, we selected not only O {\sc v} but also O {\sc vi}.
The wavelength, peak and range of temperature for these two lines are summarized in Table \ref{table1}.

We performed the following analysis for each of the two lines.
The EVE has no spatial resolution and
their original data are in Sun-as-a-star formats i.e., the data are integrated over the solar full disk.
The wavelength of each line center is affected by instrumental and orbital effects \citep{Chamberlin2016SoPh, XuETAL2022, LuETAL2023}.
Thus, we first fitted the pre-event spectrum of each selected line, which is integrated over the pre-event time of 01:39 UT to 01:59 UT, with a single Gaussian and used its center as a reference wavelength.
Subsequently, the pre-event spectrum was subtracted from each spectrum at each time.
Finally, the pre-event subtracted spectra were normalized by the peak irradiance of the pre-event spectrum. 
The continuum in EUV range is drastically weaker than the line irradiance.
Thus, we chose peak irradiance for this normalization instead of the continuum. 
In this paper, we obtained the light curves with wide wavelength integrations to include the Doppler shifted components.
The spectral integrations were performed over the central wavelength $\pm \Delta\lambda$ where $\Delta\lambda$ is $2.0$ {\AA} for O {\sc v} 629.7 {\AA} and $3.0$ {\AA} for O {\sc vi} 1031.9 {\AA} , respectively.

%------------------------------------------
\begin{deluxetable*}{ccccc}
\label{table1}
\tablenum{1}
\tablecaption{Summary of the EUV lines}
\tablewidth{0pt}
\tablehead{
\colhead{No.} &
\colhead{Ions} &
\colhead{Wavelength [{\AA}]} & 
\colhead{Peak Temperature (K)} &
\colhead{Temperature Range (K)}
}
\decimals
\startdata
(1) & O {\sc v} &
629.730 &
$10^{5.4}$ &
$10^{5.0}-10^{5.8}$\\
(2) & O {\sc vi} &
1031.90 &
$10^{5.5}$ &
$10^{5.2}-10^{6.3}$\\
\enddata
\tablecomments{Wavelengths and peak temperature are obtained from EVE documentation (\url{https://lasp.colorado.edu/eve/data_access/eve-documentation/index.html}). Temperature range is defined as the range of temperature with the ion fraction higher than $10^{-3}$, which is calculated using CHIANTI 10.1 \citep{Dere1997AAS..125..149D, Dere2023ApJS..268...52D} and ChiantiPy 0.15.1 class ChiantiPy.core.ioneq \citep{Dere2013ascl.soft08017D}.}
\end{deluxetable*}
%-------------------------------------
%\clearpage
%============================================================
\section{Results and correspondence with spatially resolved images\label{Re}}
This section presents the results of Sun-as-a-star analyses and describes their correspondence with spatially resolved images. The original images taken by SMART/SDDI were used for comparison of Sun-as-a-star H$\alpha$ spectra, while EUV images taken by SDO/AIA are used as proxies for spatially resolved images for full-disk integrated data of the SDO/EVE. Figure \ref{LC} shows the light curves of GOES soft X-ray (a), differenced H$\alpha$ equivalent width $\Delta EW$ (b), AIA 304 {\AA}, 171 {\AA}, 193 {\AA} (c), O {\sc v} 629.7 {\AA} (d), and O {\sc vi} 1031.9{\AA} (e). 
The AIA data were integrated over the solar full disk.
All light curves in Figure \ref{LC} show brightening associated with the M8.7 flare.
The detailed time evolution of these light curves is discussed along with the Sun-as-a-star spectra.

%\----------------------------------------------------------
\subsection{Result of SMART/SDDI analysis; Sun-as-a-star H$\alpha$ spectra\label{Re_Ha}}
\begin{figure}[htbp]
\centering
\includegraphics[width=12cm]
{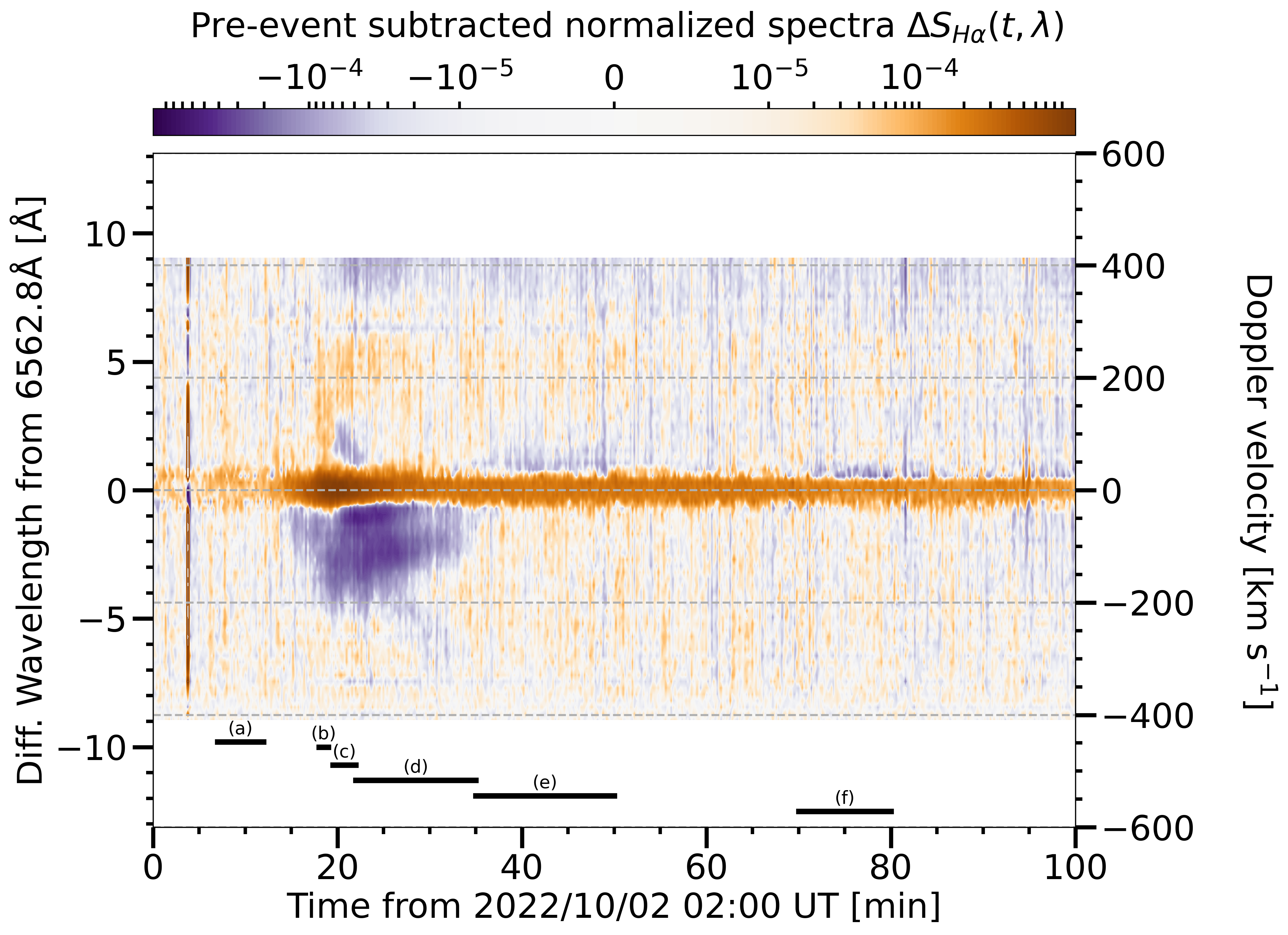}
\caption{H$\alpha$ dynamic spectrum. Spatially integrated pre-event subtracted H$\alpha$ spectra normalized by full-disk integrated continuum is shown as two dimensional color map. Positive and negative velocities mean redshifted and blueshifted ones, respectively.
Orange and purple indicate emission and absorption compared with pre-event state, respectively.
The horizontal dashed lines indicate the Doppler velocities of $+400$ km s$^{-1}$, $+200$ km s$^{-1}$, $0$ km s$^{-1}$, $-200$ km s$^{-1}$, and $-400$ km s$^{-1}$ from top to bottom. The integration time in panels (a)-(f) of Figure \ref{Ha_spectra} are indicated as horizontal black solid lines with the labels of (a)-(f), respectively.
}
\label{Ha_dynamic}
\end{figure}

\begin{figure}[htbp]
\centering
\includegraphics[width=19cm]
{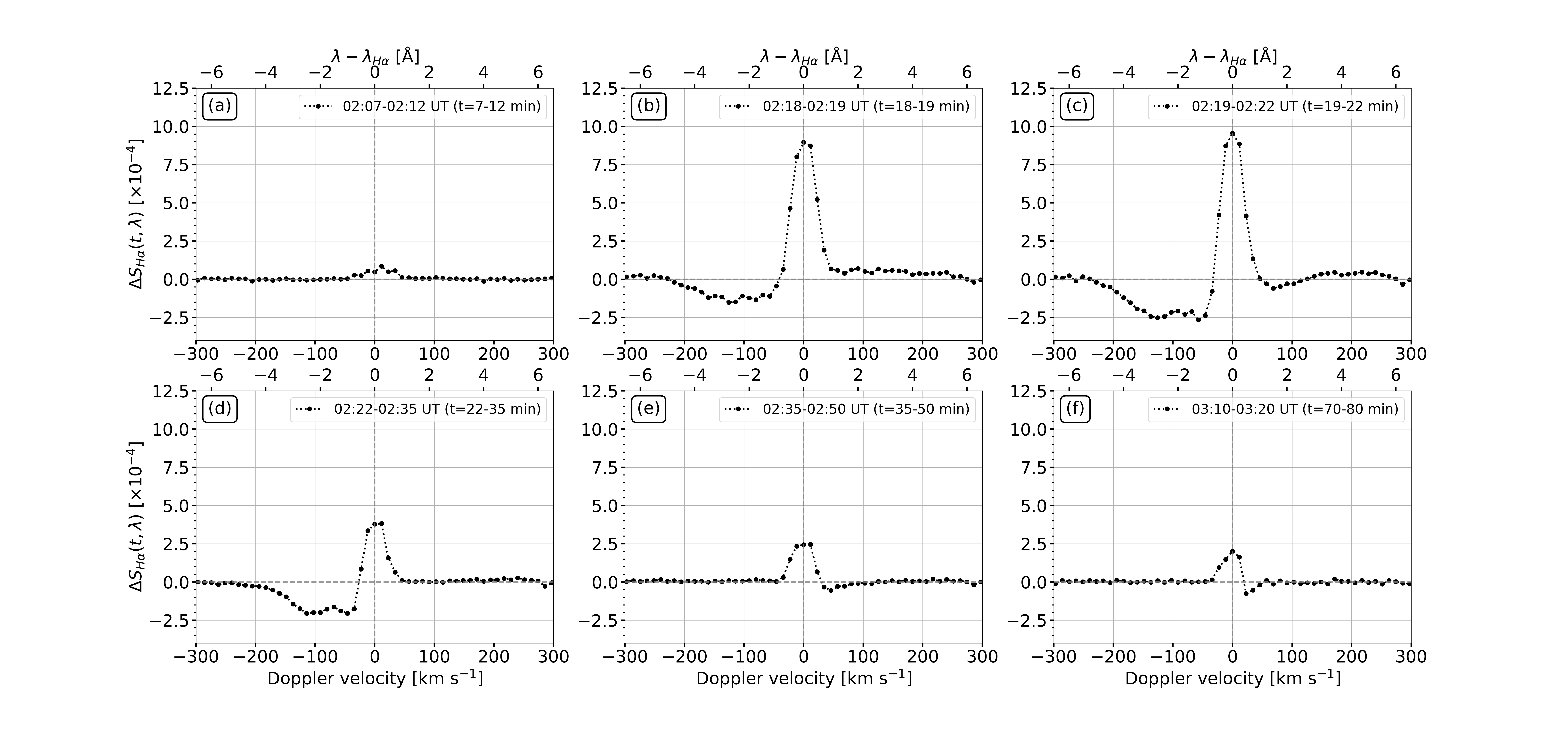}
\caption{Time series of Sun-as-a-star H$\alpha$ spectra. Pre-event subtracted H$\alpha$ spectra normalized by full-disk integrated continuum are shown. 
Each spectrum is obtained by time average in time span shown in each panel.
Note that the time spans for time average of panels (a)-(f) are indicated as black horizontal solid lines in Figure \ref{Ha_dynamic}.}
\label{Ha_spectra}
\end{figure}

Figures \ref{Ha_dynamic} and \ref{Ha_spectra} show the results of H$\alpha$ Sun-as-a-star analysis of the M8.7 flare observed on 2022 October 2. Figure \ref{Ha_dynamic} shows the dynamic spectrum of the Sun-as-a-star H$\alpha$ spectra $\Delta S_{H\alpha}(t, \lambda)$ (see Section \ref{Ha_Sun_as_a_star}), while Figure \ref{Ha_spectra} shows the Sun-as-a-star H$\alpha$ spectra averaged over some time period. The integral time span for each spectrum in Figure \ref{Ha_spectra} is indicated by the black horizontal lines in Figure \ref{Ha_dynamic}.

\begin{figure}[htbp]
\centering
\includegraphics[width=18cm]
{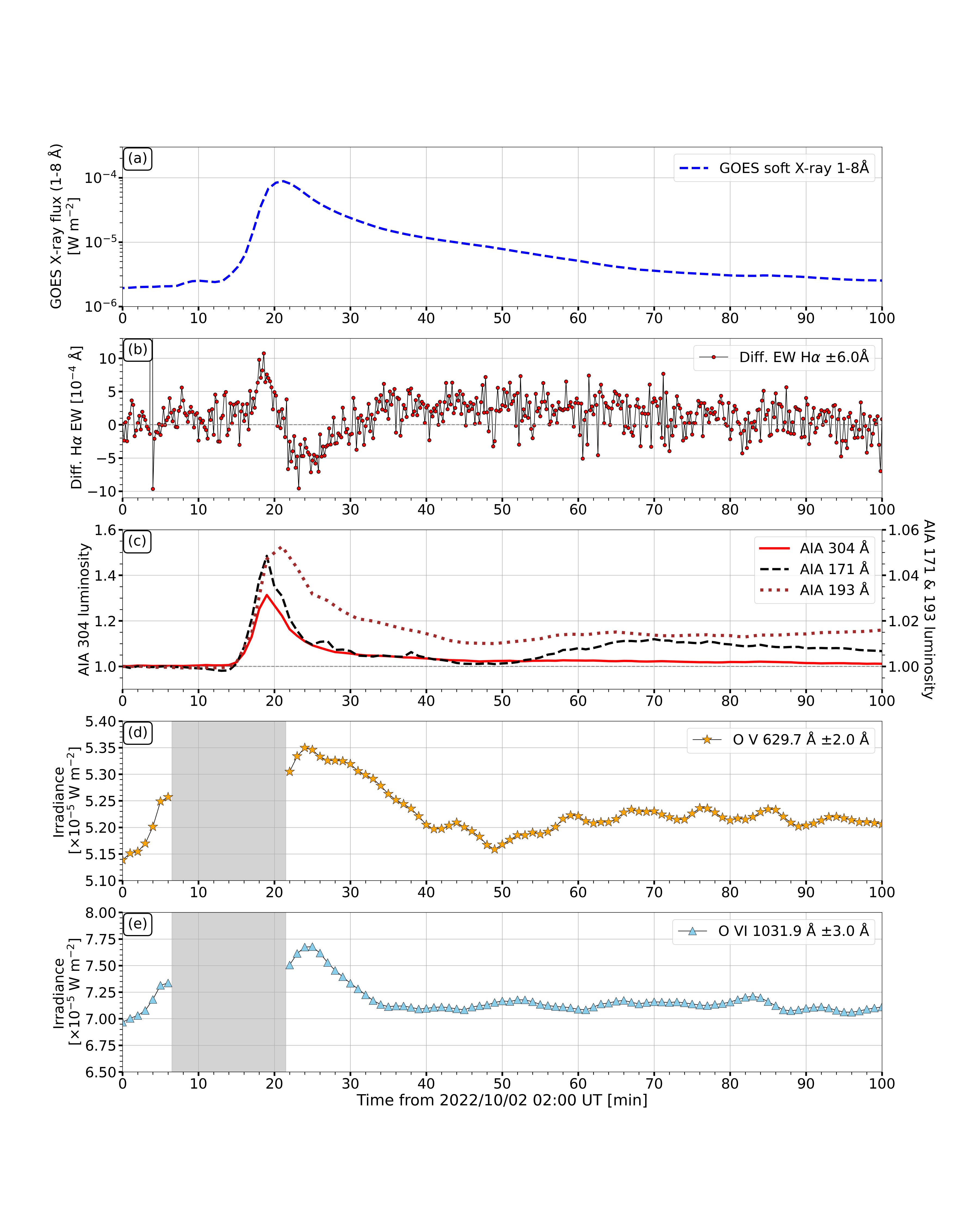}
\caption{Light curves of the M8.7 flare. From top to bottom, the light curves of GOES soft X-ray, differenced H$\alpha$ equivalent width $\Delta EW$, AIA (304, 171, 193 {\AA}), O {\small V} 629.7 {\AA}, and O {\small VI} 1031.9 {\AA} are shown in panels (a), (b), (c), (d), and (e), respectively. 
The AIA data were integrated over the solar full disk.
Note that no SDO/EVE MEGS-B data was observed in the gray region of panels (d) and (e).}
\label{LC}
\end{figure}

Corresponding to the increase of the differenced H$\alpha$ equivalent width from $t=15$ min (Figure \ref{LC} (b)), the H$\alpha$ dynamic spectrum exhibited the distinct brightening near the H$\alpha$ line center from $t=15$ min (Figure \ref{Ha_dynamic}).
This brightening near the H$\alpha$ line center mainly comes from the the flare ribbons (Figure \ref{Eruption} (b-1)-(b-3)).
Note that, the filament eruption in this event was observed as a bright feature in the H$\alpha$ line center images (Figure \ref{Eruption} (b-1)), which also contributes to the brightening near the H$\alpha$ line center in the Sun-as-a-star H$\alpha$ spectra.  
Corresponding to the red asymmetry attributed to the chromospheric condensation at the flare ribbons in the spatially resolved images (Figure \ref{RA}), the Sun-as-a-star H$\alpha$ spectra exhibited redshifted excess emission ($t=19$-$20$ min in Figure \ref{Ha_dynamic}, Figure \ref{Ha_spectra} (b)). The line broadening due to the Stark effect may also contribute to this excess emission.
Additionally, the Sun-as-a-star H$\alpha$ spectra exhibited three groups of shifted absorptions;
(i) The blueshifted absorption (t=15-35 min in Figure \ref{Ha_dynamic}) with the redshifted absorption (t=19-22 min in Figure \ref{Ha_dynamic}). (ii) The redshifted absorption after the bluesfifted absorption in (i) (t=35-50 min in Figure \ref{Ha_dynamic}). (iii) The redshifted absorption during the decay phase of the flare (t=70-80 min in Figure \ref{Ha_dynamic}). The correspondence of these results with the H$\alpha$ images is as follows.

\begin{itemize}
    \item [(i)] 
    % In this event, the filament erupted with approaching to the Earth, thus, blueshifted absorption clearly appeared in the dynamic spectra. 
    In this event, blueshifted absorption was clearly visible in the H$\alpha$ dynamic spectra (up to $-300$ km s$^{-1}$) which corresponds to the erupting filament (Figure \ref{Eruption} (c-1) and (d-2)).
    Additionally, as shown in the spatially resolved images of Figure \ref{Eruption} (a-1), some part of the erupting filament had redshifted components which may reflect the expanding/helical motions of the filament. 
    As a result, dynamic spectra showed not only the blueshifted absorption (t=15-35 min in Figure \ref{Ha_dynamic}, Figure \ref{Ha_spectra} (b), (c) and (d)) but also the redshifted absorption during the filament eruption (t=19-22 min in Figure \ref{Ha_dynamic}, Figure \ref{Ha_spectra} (c)).
    
    \item[(ii)] Once the blueshifted absorption almost disappeared, the redshifted absorption appeared again (up to 100 km s$^{-1}$, t=35-50 min in Figure \ref{Ha_dynamic} and Figure \ref{Ha_spectra} (e)).
    This absorption originates from the partial drainage (falling plasma) of the erupted filament (Figure \ref{Eruption} (a-3)).

    \item[(iii)] The redshifted absorption also appeared during the decay phase of the flare (up to 50 km s$^{-1}$, t=70-80 min in Figure \ref{Ha_dynamic}, Figure \ref{Ha_spectra} (f)).
    Corresponding to this resdshifted absorption, downflows can be confirmed in the SDDI H$\alpha$ red wing images (Figure \ref{CRs} (c-1) and (c-2)). As introduced in Section \ref{PFL_Dim}, the region of the downflows matches that of post-flare loops in AIA 171 {\AA} and 193 {\AA} images (Figure \ref{CRs} (a-1) and (b-1)). This agreement suggests that cool materials flow along the post-flare loops towards solar surface.
    Thus, this redshifted absorption in the Sun-as-a-star H$\alpha$ spectra could be attributed to the cool downflows along the post-flare loops.
     
\end{itemize}

As described above, the Sun-as-a-star H$\alpha$ spectra in this study reflect the three key processes in the flare standard model, i.e., filament eruption, chromospheric condensation, and cool downflows along the post-flare loops. While each one of these processes was individually identified in previous Sun-as-a-star analyses \citep[e.g.,][]{OtsuETAL2022}, this Sun-as-a-star result for the first time, presents the signatures related to all three processes within a single flare.

The H$\alpha$ equivalent width started to increase at almost the same time as the soft X-ray, and reached its peak time (t= 18 min in Figure \ref{LC} (b)) before the GOES peak time (t= 21 min in Figure \ref{LC} (a)), which has been confirmed in the Sun-as-a-star analyses of the other flares \citep{NamekataETAL2022c, OtsuETAL2022}.
Moreover, because of the absorption coming from the filament eruption, the H$\alpha$ equivalent width deviated from the soft X-ray (especially at t=20-30 min in Figure \ref{LC} (a) and (b)). 
This deviation of the H$\alpha$ from the soft X-ray has also been previously reported \citep[e.g., ][]{NamekataETAL2022a,OtsuETAL2022}.
The redshifted absorptions due to (ii) the falling plasma and (iii) the cool downflows along the post-flare loops--which were fainter and slower than (i) the shifted absorptions due to the filament eruption--could not be confirmed in the H$\alpha$ equivalent width. 
%------------------------------------------------------------
\subsection{Results of SDO/EVE and AIA analysis; Sun-as-a-star O {\sc v} 629.7 {\AA} and O {\sc vi} 1031.9 {\AA} spectra \label{Re_EUV}}
Here we show the spatially integrated spectra of O {\sc v} 629.7 {\AA} and O {\sc vi} 1031.9 {\AA} lines.
Note that the data of EVE MEGS-B are missing from 02:07 UT to 02:20 UT when the flare showed impulsive increase in GOES soft X-ray\footnote{According to EVE science operations mission log, EVE worked under normal operations on 2022 October 2 (\url{https://lasp.colorado.edu/eve/data_access/eve_data/EVE_sciopslog.html}). }(Figure \ref{LC} (d) and (e)).

\subsubsection{O {\sc v} 629.7 {\AA}\label{Re_Ov}}
\begin{figure}[htbp]
\centering
\includegraphics[width=15cm]
{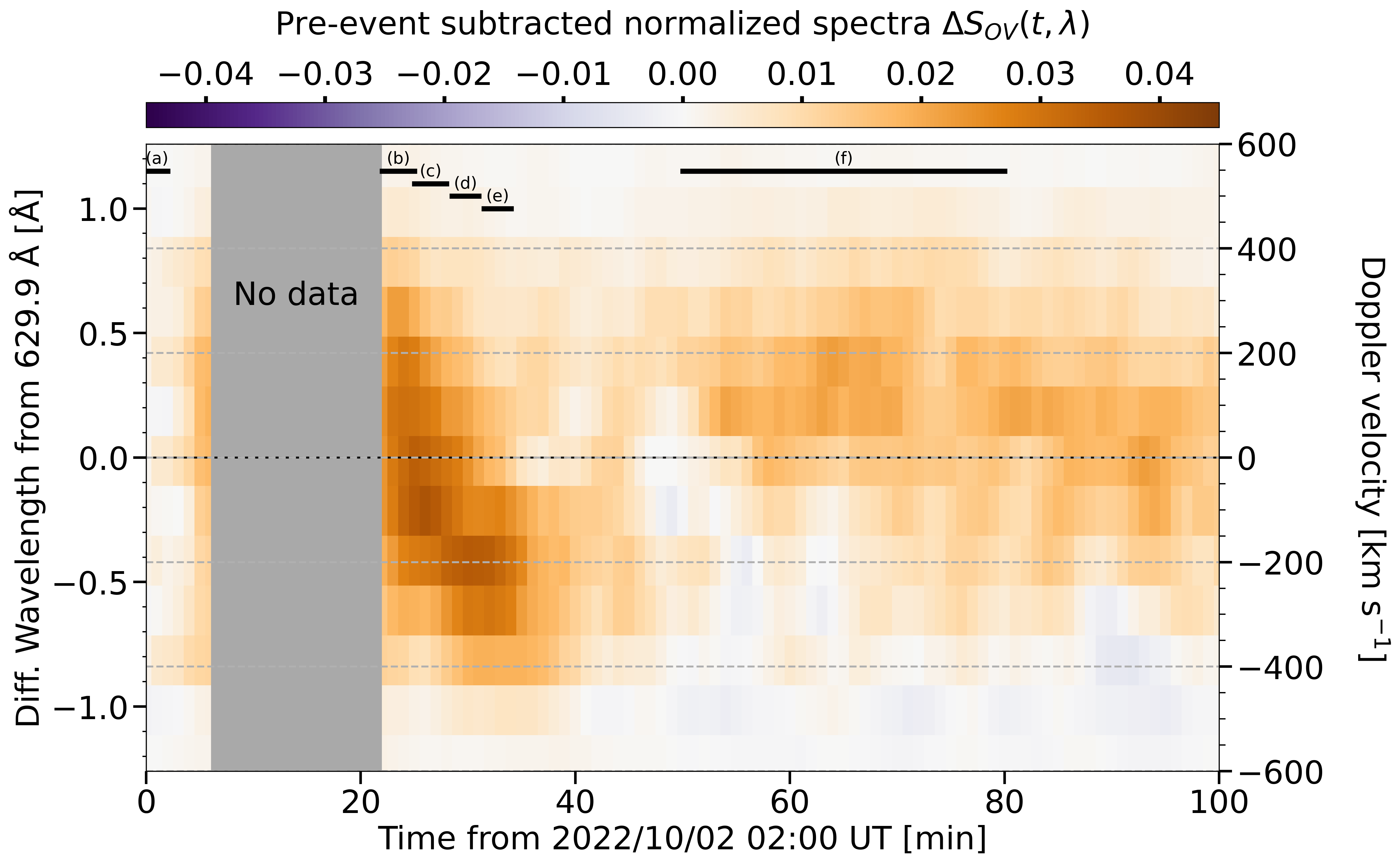}
\caption{The same as Figure \ref{Ha_dynamic}, but for O {\small V} 629.7 {\AA}. 
The spectra are normalized by The peak irradiance of the pre-event spectrum obtained by Gaussian fitting.
Note that no SDO/EVE MEGS-B data was observed in the gray region.
}
\label{OV_dynamic}
\end{figure}

\begin{figure}[htbp]
\centering
\includegraphics[width=19cm]
{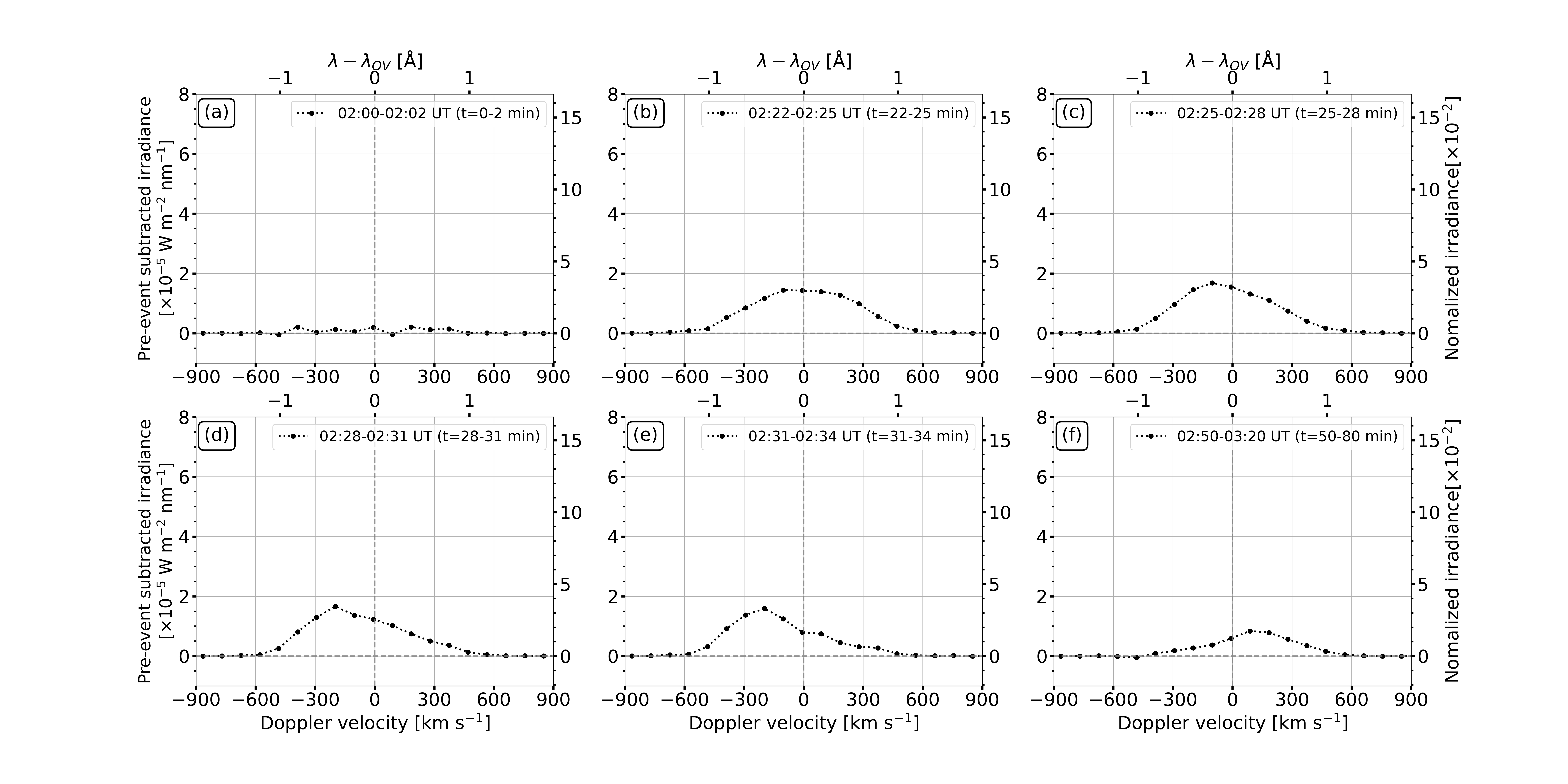}
\caption{Time series of O {\small V} spectra. Pre-event subtracted O {\small V} spectra are shown. The left and right vertical axes correspond to irradiance in the real scale and normalized irradiance. The normalization is the same for Figure \ref{OV_dynamic}.}
\label{OV_spectra}
\end{figure}

Figures \ref{OV_dynamic} and \ref{OV_spectra} show the results for O {\sc v} 629.7 {\AA}, similar to the results of H$\alpha$ in Figures \ref{Ha_dynamic} and \ref{Ha_spectra}. The O {\sc v} dynamic spectrum in Figure \ref{OV_dynamic} exhibits two main features: (i) Blueshifted and redshifted brightening around the peak time of GOES soft X-ray flux ($t\approx20-40$ min in Figure \ref{OV_dynamic}). (ii) The Second redshifted brightening during the flare decay phase (after $t\approx50$ min). 
The comparisons of these results with AIA images are as follows. 

\begin{itemize}
    \item [(i)] The O {\sc v} spectra exhibited blueshifted brightening with drifting to higher velocity i.e. accelerating (up to $-400$ km s$^{-1}$, t=22-40 min in Figure \ref{OV_dynamic} and \ref{OV_spectra} (b)-(e)).
    Meanwhile, AIA 171 {\AA} and 304 {\AA} images demonstrated the filament eruption as bright features compared to the solar surface (Figure \ref{Eruption} (e-1)-(e-3), (f-1)-(f-3)). 
    This time correspondence suggests that the blueshifted brightening in O {\sc v} originates from the filament eruption.
    Additionally, AIA 171 {\AA} and 304 {\AA} have typical temperatures of $\sim10^{6}$ K and $\sim10^{4.9}$ K, respectively. The peak temperature of O {\sc v} 629.7 {\AA} line ($\sim10^{5.4}$ K) is consistently located within them.
    Note that the typical temperature of AIA 193 {\AA} is higher than $>10^{6}$ K, and their images mainly show the eruption as a dark feature (Figure \ref{Eruption} (g-2)-(g-3)), implying that the temperature of the erupting material was typically lower than $10^{6}$ K.
    Moreover, the dynamic spectra also showed redshifted brightening around the flare impulsive phase.
    This could be attributed to the expanding/helical motion of the filament, similar to the case of the redshifted absorption accompanying the blueshifted absorption in the Sun-as-a-star H$\alpha$ spectra (t=19-22 min in Figure \ref{Ha_dynamic}).
    However, the flare-related downflows such as chromospheric condensation can also cause this redshifted brightening. 
    This redshifted component has not been focused in this paper because of the lack of the EVE data around the flare impulsive phase.
    \item[(ii)] During the decay phase of the flare, the redshifted brightening was confirmed again (after t=50 min in Figure \ref{OV_dynamic}, Figure \ref{OV_spectra} (f)).
    Correspondingly, the secondary increase appeared in the light curve of O {\sc v} 629.7 {\AA} from t=50 min (Figure \ref{LC} (d)). Furthermore, the light curves of AIA 193 {\AA} and 171 {\AA} show the secondary increase from t=50 min (Figure \ref{LC} (c)). Simultaneously, AIA 193 {\AA} and 171 {\AA} images demonstrated the growing post-flare loops (Figure \ref{CRs} (a-1)-(a-2) and (b-1)-(b-2)), which suggests that this redshifted brightening in O {\sc v} is related to the post-flare loops.
\end{itemize}

\subsubsection{O {\sc vi} 1031.9 {\AA} \label{Re_OVI}}
\begin{figure}[htbp]
\centering
\includegraphics[width=15cm]
{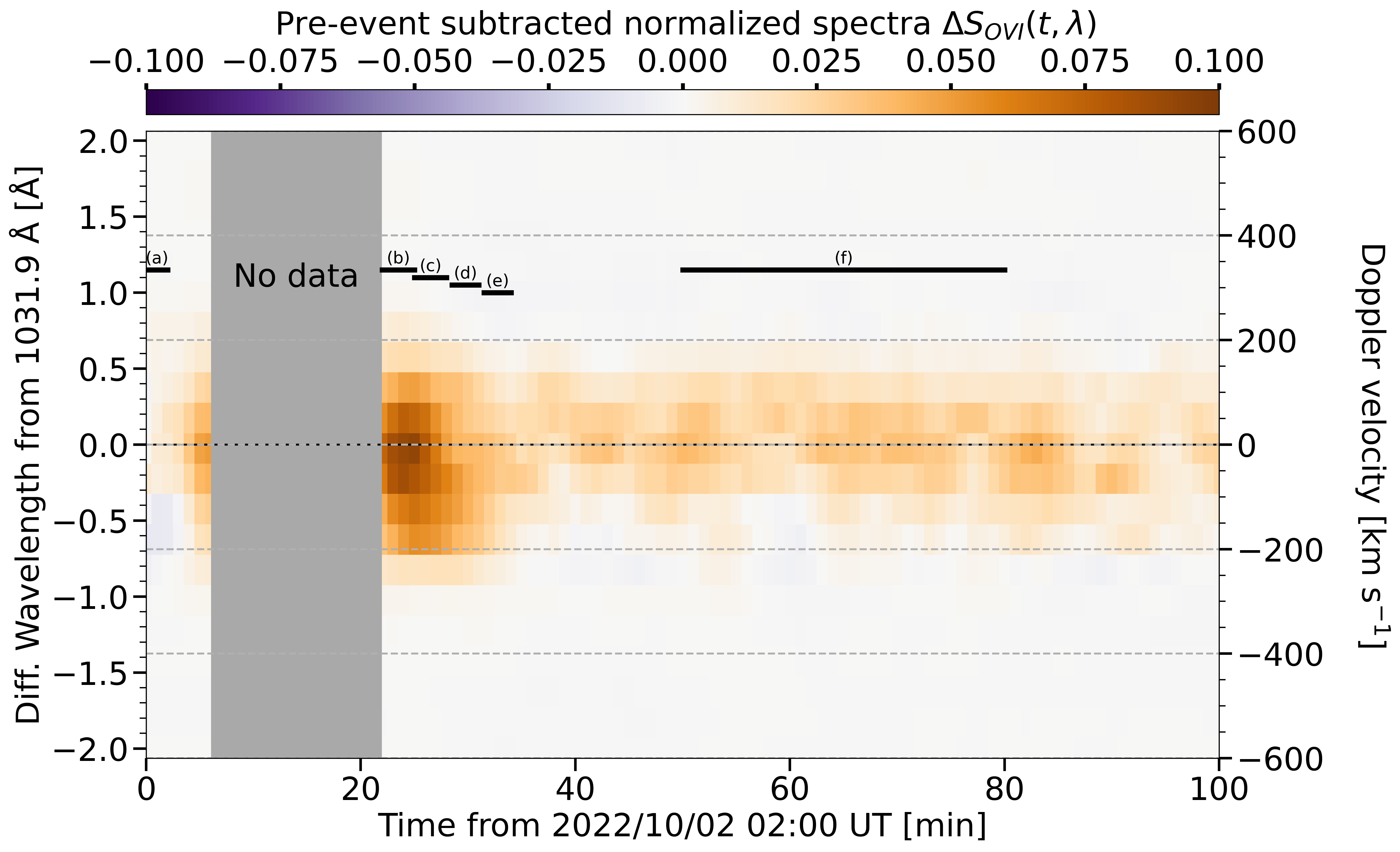}
\caption{The same as Figure \ref{OV_dynamic}, but for O {\small VI} 1031.9 {\AA}. 
Note that no SDO/EVE MEGS-B data was observed in the gray region}
\label{OVI_dynamic}
\end{figure}

\begin{figure}[htbp]
\centering
\includegraphics[width=19cm]
{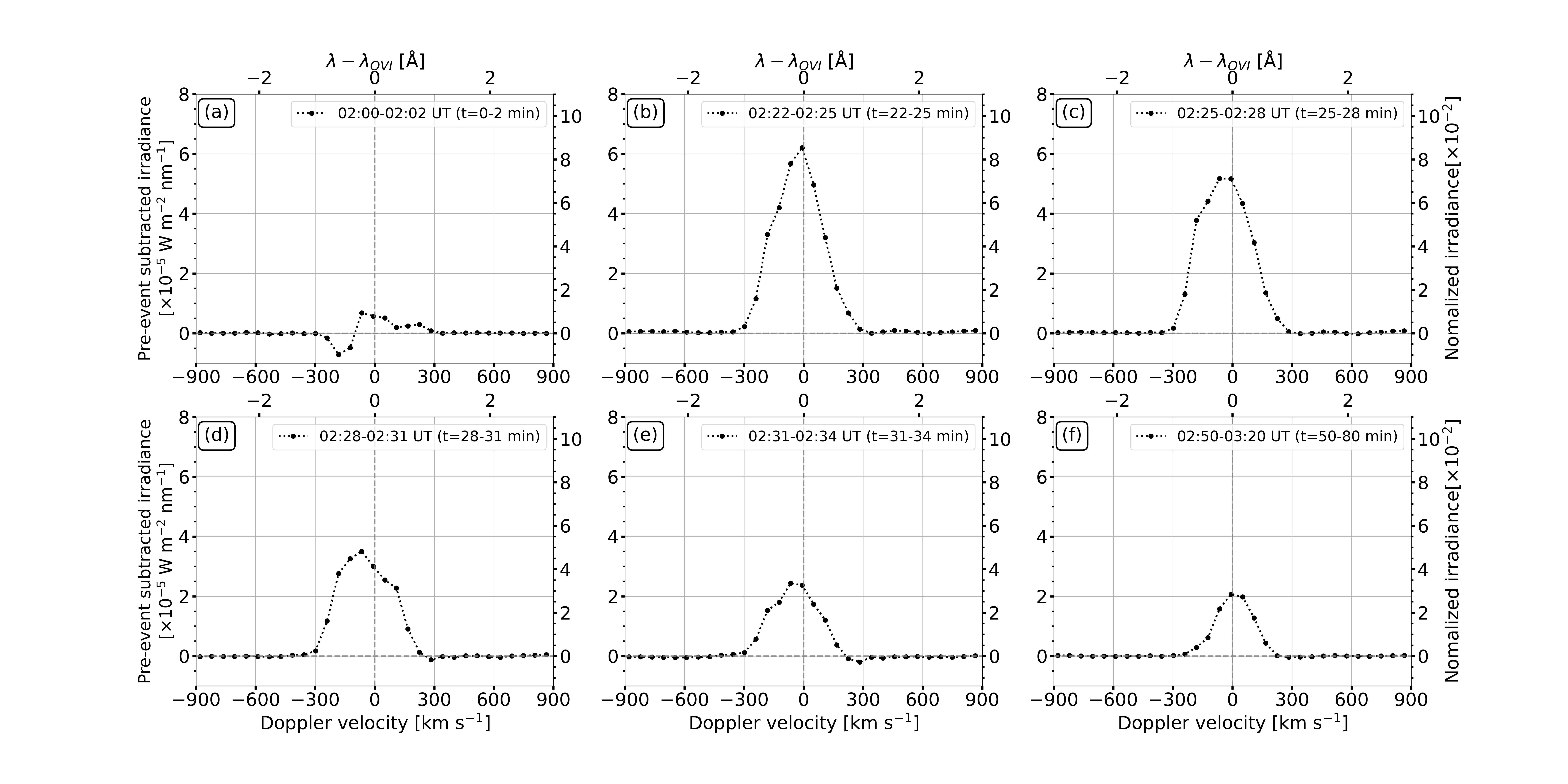}
\caption{The same as Figure \ref{OV_spectra}, but for O {\small VI} 1031.9 {\AA}. Note that the scale of the left vertical is the same as that in Figure \ref{OV_spectra}.}
\label{OVI_spectra}
\end{figure}

Figures \ref{OVI_dynamic} and \ref{OVI_spectra} show the results for O {\sc vi} 1031.9 {\AA}.
Although O {\sc vi} 1031.9 {\AA} shows blueshifted brightening (Figure \ref{OVI_spectra} (b)-(e)), its maximum speed is approximately $-200$ km s$^{-1}$, which are slower than that in O {\sc v} 629.7 {\AA} (around $-400$ km s$^{-1}$). 
This blueshifted brightening possibly corresponds to the filament eruption, similar to that in O {\sc v} 629.7 {\AA}. 
The spectra of O {\sc vi} did not exhibit redshifted brightening during the flare decay phase.
Correspondingly, the light curve of O {\sc vi} showed no clear secondary increase unlike O {\sc v}.

% The reason for difference between O {\small V} 629 {\AA} and O {\small VI} 1031 {\AA} is discussed in Section \ref{Dis}.   
%------------------------------------------------------------
\subsection{Composite dynamic spectrum of H$\alpha$ and O {\sc v} 629.7 {\AA} \label{Re_Ha_OV}}
\begin{figure}[htbp]
\centering
\includegraphics[width=15cm]
{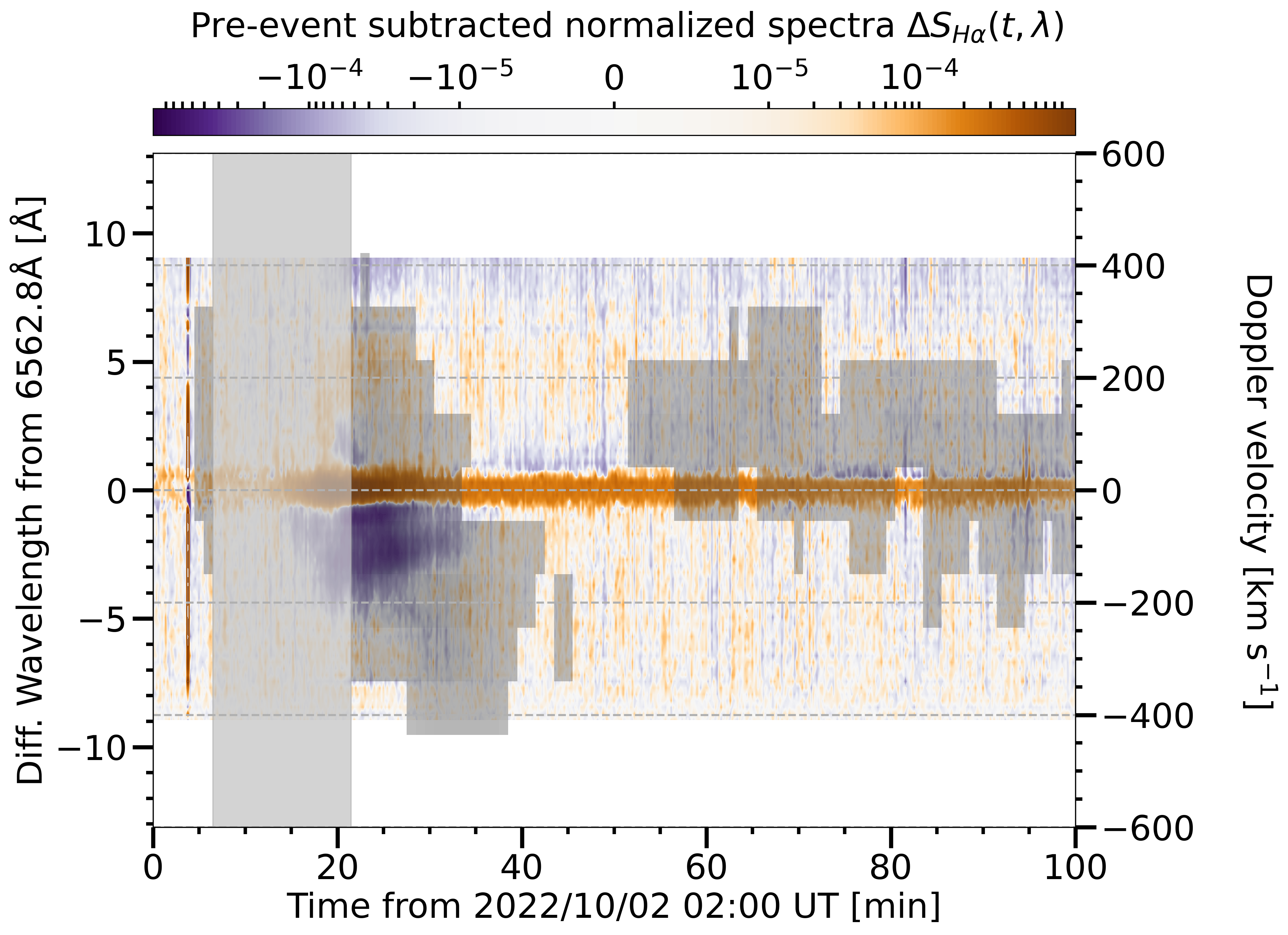}
\caption{The composite dynamic spectra of H$\alpha$ and O {\small V} 629.7 {\AA}. Whereas the H$\alpha$ dynamic spectrum is same as that in Figure \ref{Ha_dynamic}, the O V spectra with the the absolute change of irradiance larger than the threshold (i.e. bright components in Figure \ref{OV_dynamic}) are shown as black color (see Section \ref{Re_Ha_OV} for the details of the threshold).}
\label{Ha_OV}
\end{figure}

Figure \ref{Ha_OV} shows the composite dynamic spectra of H$\alpha$ and O {\sc v} 629.7 {\AA}. The H$\alpha$ dynamic spectrum is the same as that in Figure \ref{Ha_dynamic}.
The dynamic spectrum for O {\sc v} 629.7 {\AA} is plotted as black color only with the absolute change of irradiance larger than 3$\sigma_{max}$ where $\sigma_{max}$ is the maximum value of the time-averaged standard deviation with respect to the wavelength calculated using the pre-event data. In this event, $\sigma_{max}$ for the normalized irradiance (see Section \ref{EUV_SAS}) is approximately equal to $4.15\times10^{-3}$, which is about 10 \% of the maximum brightening ($\sim4\times10^{-2}$) in O {\sc v} 629.7 {\AA} (Figure \ref{OV_spectra}).

The blueshifted absorption in H$\alpha$ and blueshifted brightening in O {\sc v} 629.7 {\AA} are initially matched well (slower than $-200$ km in Figue \ref{Ha_OV}).
Although H$\alpha$ blueshifted absorption became almost invisible faster than $-200$ km s$^{-1}$, the blueshifted brightening was still clear in O {\sc v} 629.7 {\AA} (at least brighter than $3\sigma_{max}$) and its velocity reached to $-400$ km s$^{-1}$.
The difference in visibility between H$\alpha$ and EUV can be also confirmed in spatially resolved images of the SDDI and AIA images as follows.
Even when the erupting materials became almost invisible in H$\alpha$ at around t=34 min (Figure \ref{Eruption} (c-3), (d-3)),
the AIA 304 {\AA} and 171 {\AA} images still exhibited the erupting materials as bright features (Figure \ref{Eruption} (e-3) and (f-3)).

%=============================================================
\section{Discussion \label{Dis}}
%-------------------------------------------------------------
%\subsection{Filament eruption in Sun-as-a-star spectra}
\subsection{New points in the Sun-as-a-star H$\alpha$ spectra}
In the Sun-as-a-star H$\alpha$ spectra of this event, chromospheric condensation , the filament eruption, and cool downflows along the post-flare loops appeared as redshifted excess emission during the flare impulsive phase, dominant blueshifted absorption during the flare impulsive phase, and redshifted absorption during the flare decay phase, respectively.
Herein, the signatures related to all three processes have been presented within a single flare for the first time, although each one of these processes has been individually reported in the prior Sun-as-a-star analyses \citep{NamekataETAL2022a,NamekataETAL2022c,OtsuETAL2022}. 
This result emphasizes that H$\alpha$ spectra enable us to grasp the whole picture of flares from impulsive phases (chromospheric condensation and filament eruptions) to decay phases (cool downflows along post-flare loops), even via spatially integrated spectra.
However, the redshifted absorption and redshifted excess emission (and/or line broadening) during the impulsive phase cancelled each other. Such cancellation in Sun-as-a-star H$\alpha$ spectra was also reported by \citet{OtsuETAL2022}, and it may hinder us from accurately deducing the active phenomena from spatially integrated data.
To investigate the dynamics of cool plasma ($\sim10^{4}$ K) without the cancellation between absorption and emission, performing simultaneous observations of multiple chromospheric lines may be useful (e.g., Ca II H\&K and He I 10830 {\AA}) because of the varieties of their sensitivity to the slight difference in plasma parameters (e.g., temperature and density).
%---------------------------------------------
\subsection{Filament eruption in distinctly different temperature lines : H$\alpha$ and O {\sc v} 629.7 {\AA}}
Here, we focus on H$\alpha$ and O {\sc v}, which exhibited clear Doppler signatures.
As shown in Sections \ref{Re_Ha}, \ref{Re_Ov} and \ref{Re_Ha_OV}, the filament eruption was simultaneously observed in H$\alpha$ and O {\sc v} spectra as dominant blueshifted component.
At the initial phase of the eruption, both exhibited the shifted component simultaneously.
H$\alpha$ and O {\sc v} are formed at the temperatures of $\sim10^{4}$ K and $\sim10^{5.4}$ K, respectively. Therefore, the simultaneous detection of the eruption in these two lines means that the filament had multi-temperature structure during the early phase of the eruption. 
The erupting filament is expected to form 'shell' structure, which has a cooler core of $\sim10^{4}$ K and surrounding hotter shells of $\sim10^{5}-10^{6}$ K  \citep[e.g.,][]{RiveraETAL2019ApJS}.
Such a shell structure for a solar eruption has also been reconstructed in noble realistic magnetohydrodynamic simulations \citep[e.g.,][]{Chen2023ApJ...950L...3C}. 
In spatially resolved images, the eruption was observed with smaller spatial scale in H$\alpha$ (Figure \ref{Eruption} (a)-(d)), corresponding to the cool core of the eruption.
Meanwhile, EUV images showed the eruption with larger spatial scale compared with that in H$\alpha$ (Figure \ref{Eruption} (e)-(g)), which would correspond to  the surrounding hotter shells. In addition to the spatially resolved observation of multi-temperature structure, the present study demonstrated that the multi-temperature structure of the filament eruption can be detected even in spatially integrated spectra for the first time.
Recent stellar spectroscopic observations have provided probable detections of stellar filament/prominence eruptions \citep[e.g.,][]{MaeharaETAL2021,NamekataETAL2022a,InoueETAL2023ApJ};
however, their temperature structures have not been investigated in detail.
Our Sun-as-a-star analysis results suggest that multi-temperature structures of stellar eruptions can be observed by combining H$\alpha$ and EUV lines, even in spatially integrated spectra.

At the late phase of the eruption, while the blueshifted absorption in H$\alpha$ became fainter, O {\sc v} brightening remained clear and reached a blueshifted velocity of $\sim-400$ km s$^{-1}$.
The erupting filament was heated up while traveling through the hot corona, and the core may become hotter. 
Thus, it is natural that the filament eruption becomes fainter in H$\alpha$ spectra at the late phase of the eruption as often observed in H$\alpha$ line observations of filament eruptions \citep[e.g.,][]{OtsuETAL2022}.
Furthermore, O {\sc v} has about one order higher formation temperature ($\sim10^{5.4}$ K) than H$\alpha$ ($\sim10^{4}$ K).
Due to this difference of the formation temperature, the eruption remained visible for a longer time in O {\sc v} compared with H$\alpha$.
The changes of observable lines from lower to higher temperatures have been reported in previous studies \citep[e.g.,][]{Fontenla1989,Chifor2006A&A...458..965C}, but the present study, for the first time, demonstrated that such changes can be observed in spatially integrated solar spectra.
Our Sun-as-a-star results suggest that EUV spectra can enable the tracking of stellar filament eruptions for a longer time than H$\alpha$ even in spatially integrated spectra. 
In stellar cases, many of Doppler shifted components in H$\alpha$ spectra expected to originate from stellar filament eruptions become invisible before their velocities reach the escape velocities at the stellar surfaces \citep[e.g.,][]{Vida2019A&A...623A..49V,MuhekiETAL2020,MaeharaETAL2021,NamekataETAL2022a}. However, our result indicates that the erupting materials may still be present and observable in EUV spectra, even when they become almost invisible in the spatially integrated H$\alpha$ spectra.

%----------------------------------------
\subsection{Visibility of filament eruption in different EUV lines : O {\sc v} 629.7 {\AA} and O {\sc vi} 1031.9 {\AA}}
As shown in Section \ref{Re_EUV}, O {\sc vi} demonstrated relatively lower velocity for blueshifted component ($\sim-200$ km s$^{-1}$) unlike O {\sc v}, exhibiting the velocity of $\sim-400$ km s$^{-1}$.
O {\sc v} and O {\sc vi} have the peak formation temperatures of $10^{5.4}$ K and $10^{5.5}$ K, respectively. However, they have wide temperature range than just the peak, in particular, O {\sc vi} has wider range and the hotter edge of O {\sc vi} is larger than $10^{6}$ K whereas that of O {\sc v} is lower than $10^{6}$ K (Table \ref{table1}).
Thus, in a simple explanation, the erupting material with high velocity is not heated sufficiently to become high-temperature plasma emitting in O {\sc vi} line.
The scarcity of high-temperature plasma around $10^{6}$ K can also be confirmed in the AIA 193 {\AA} image (Figure \ref{Eruption} (g-3)), where most of the erupting materials were observed as dark features.
Considering the result for H$\alpha$, the cooler materials ($10^{4}$ K) were heated adequately to become plasmas with O {\sc v} 629.7 {\AA} temperature (higher than $10^{5}$ K but lower than $10^{6}$ K, Table \ref{table1}) and these materials remained stable at this temperature. 
If this kind of eruptions occur on a distant star, observing O {\sc v} is better to capture fast components of the eruption compared with O {\sc vi}.
In reality, O {\sc vi} 1031.9 {\AA} can avoid strong absorption by interstellar medium; thus, this result is somewhat negative from a stellar observational standpoint.
However, it is expected that the temperature of erupting materials as a result of heating are different in each event and there are various cases of filament eruption with respect to the thermal state of materials. 
In the future, we need to statistically investigate which EUV spectral lines are best for detecting each case of filament eruptions using solar data, for stellar observation.
%-------------------------------------------------------------
\subsection{The decay phase of the flare in O {\sc v} 629.7 {\AA}: Redshifted brightening and dimming}
In the decay phase of the flare, the O {\sc v} spectra showed redshifted brightening (up to $200$ km s$^{-1}$, Figure \ref{OV_dynamic} and Figure \ref{OV_spectra} (f)).
To the best of our knowledge, this is the first report on the redshifted brightening during the decay phase of the flare in O {\sc v} spectra taken by SDO/EVE.
As mentioned in Section \ref{Re_Ov}, this redshifted brightening may be related to the post-flare loops because the secondary increase of the O {\sc v} light curve is similar to those of AIA 193 {\AA} and 171 {\AA} (especially, 171 {\AA}), which is ascribed to the evolving post-flare loops.
Similar to the redshifted absorption in the Sun-as-a-star H$\alpha$ spectra during the decay phase of the flare (Section \ref{Re_Ha} (iii)), this redshifted brightening in the O {\sc v} spectra probably originates from downflows toward the solar surface along the post-flare loops.
It should be noted that the redshifted absorption during the decay phase in H$\alpha$ began to appear slightly after the redshifted brightening in O {\sc v} (Figure \ref{Ha_OV}).
Cooling of post-flare loops provides the downflows in cooler lines (e.g., H$\alpha$) after those in relatively hotter lines (EUV), thus it is natural for the appearance of the H$\alpha$ downflows delay to the O {\sc v} (EUV) downflows.
Meanwhile, O {\sc v} redshifted brightening has the maximum Doppler velocity of $\sim 200$ km s$^{-1}$, which is faster than that of the H$\alpha$ redshifted absorption during the decay phase ($\sim 50$ km s$^{-1}$). Some theoretical studies have shown that the velocity of downflows along the loop structure is affected by difference in plasma parameters, such as density and pressure \citep[e.g.,][]{Oliver2014ApJ...784...21O}.
Thus, the different velocities of the redshifted components in H$\alpha$ and EUV can be explained by the change in the plasma state in different cooling stages.

In terms of decay phases of flares, previous studies have reported that some transition region lines including O {\sc v} 629.7 {\AA} exhibited decrease in irradiance after the flare impulsive phase \citep{Harrison2003A&A...400.1071H,XuETAL2022}, which can be interpreted as being associated with CMEs.
However, the light curves of O {\sc v} in the present M8.7 flare showed no dimming (Figure \ref{LC} (d)), although the related CME was observed (Figure \ref{CME}).
Meanwhile the spatially resolved images of AIA 193 {\AA} and 171 {\AA} showed dimming (Figure \ref{CRs} (a-2), (b-2)), the full-disk integrated light curves of these two channels also did not exhibit dimming (Figure \ref{LC} (c)).
As discussed above, the post-flare loops were dominantly developing during the decay phase of the flare (Figure \ref{CRs} (a-1)-(a-2) and (b-1)-(b-2)). The dimming in AIA channels was obscured by the emission from the post-flare loops, and the AIA 193 {\AA} and 171 {\AA} full-disk integrated light curves showed no dimming.
Although the O {\sc v} in our event may have been associated with the dimming, similar to those in AIA 193 and 171 images, the contribution from the redshifted brightening, which may be related to the post-flare loops, was dominant at least in the full-disk integrated light curve.

In the case of the Sun, post-flare loops and coronal dimmings can be spatially separated (e.g., AIA observations).
However, in the case of stellar observations, they could cancel each other out, as observed in the present Sun-as-a-star observation.
The redshifted absorptions/emissions during the decay phases of stellar flares have been reported in H$\alpha$ line and they are considered as candidates for downflows along post-flare loops \citep{HondaETAL2018, NamizakiETAL2023ApJ}.
Additionally, dimmings during the decay phases of stellar flares have been recently reported and their application for detecting of stellar CMEs has been discussed \citep{VeronigETAL2021}.
Hence, stellar post-flare loops and dimmings are receiving more attention from stellar community.
It is important to clarify the condition that determines whether post-flare loops or dimmings become dominant for each spectral line utilizing spatially resolved solar data, because they are expected to become dominant during the same phase even in stellar cases.
%=============================================================
\section{Summary \label{Sum}}
We conducted the Sun-as-a-star analysis of the M8.7 flare and the associated filament eruption followed by the CME on 2022 October 2 using H$\alpha$ ($\sim10^{4}$ K) spectral images and full-disk integrated O {\sc v} 629.7 {\AA} ($\sim10^{5.4}$ K) and O {\sc vi} 1031.9 {\AA} ($\sim10^{5.5}$ K) spectra observed by SMART/SDDI and SDO/EVE, respectively.
Herein, such spectroscopic Sun-as-a-star analysis simultaneously using H$\alpha$ and EUV spectra has been performed for the first time.
As a result, we found that Sun-as-a-star H$\alpha$ and O {\sc v} spectra showed blueshifted absorption and brightening, respectively.
These blueshifted components can be attributed to the filament eruption and they appeared simultaneously during the early phase of the eruption.
Subsequently, even when the blueshifted absorption became almost invisible in the Sun-as-a-star H$\alpha$ spectra, the O {\sc v} blueshifted brightening up to $-400$ km s$^{-1}$ was still clearly visible. 
This Sun-as-a-star result suggests that EUV spectra
can enable the tracking of stellar filament eruptions for a longer time than H$\alpha$, even in spatially integrated observational data.
Many of Doppler shifted components in stellar H$\alpha$ spectra expected to originate from stellar filament eruptions become invisible before their velocities reach the escape velocities at the stellar surfaces. However, our result indicates that the erupting materials may still be present and observable in EUV spectra, even when they become almost invisible in the spatially integrated H$\alpha$ spectra.
Meanwhile, O {\sc vi} spectra exhibited relatively lower speed component ($\sim-200$ km s$^{-1}$), implying that the heating was not enough to generate high-temperature plasma around $10^{6}$ K with high velocity in this event. We also found the redshifted brightening in O {\sc v} spectra during the decay phase of the flare, which can be attributed to the downflows along the post-flare loops. Although the O {\sc v} in our event may have been associated with the dimming in transition region temperature line previously reported by \citet{XuETAL2022}, the contribution from the redshifted brightening was dominant and no dimming appeared in the O {\sc v} full-disk integrated light curve even if the dimming occurred in the temperature of O {\sc v}.
Although only one event has been analyzed in this paper, various events should be analyzed in the near future to identify more common characteristics. Additionally, for the future practical applications to EUV and FUV observations of stellar activities such as the \textit{Extreme-ultraviolet Stellar Characterization for Atmospheric
Physics and Evolution} \citep[\textit{ESCAPE};][]{France2022JATIS...8a4006F}, we have to investigate which EUV or FUV lines are suitable for detecting active phenomena in each of various thermal structures using solar data.

\begin{acknowledgments}
The authors thank the anonymous referee for constructive comments that significantly improved the quality of this paper. We express our sincere gratitude to the staff of Hida Observatory for developing and maintaining the instrument and daily observation. 
We would like to acknowledge the data use from GOES, SDO, and SOHO.
SDO is a mission for NASA’s Living With a Star program. CHIANTI is a collaborative project involving George Mason University, the University of Michigan (USA), University of Cambridge (UK) and NASA Goddard Space Flight Center (USA).
This work was supported by JSPS KAKENHI grant numbers 21H01131 (A.A.). This work was also supported by JST, the establishment of university fellowships towards the creation of science technology innovation, Grant Number JPMJFS2123 (T.O.).
\end{acknowledgments}

\software{astropy \citep{Astropy}, sunpy \citep{Sunpy2020ApJ}, ChiantiPy \citep{Dere2013ascl.soft08017D}}

%=============================================================
%Figures
\clearpage

%\clearpage

%\clearpage

%\clearpage

% %results------------------------------------------------------

%\clearpage

%==============================================================
\clearpage
%\bibliography{sample631}{}

\end{document}